\documentstyle[floats,10pt,epsf,prl,aps]{revtex}
\begin{document}
\title{Dynamics of a vortex in a trapped Bose-Einstein condensate}
\author{Anatoly A.~Svidzinsky and Alexander L.~Fetter}
\address{Department of Physics, Stanford University, Stanford, CA 94305-4060}
\date{\today }
\maketitle

\begin{abstract}
We consider a large condensate in a rotating anisotropic harmonic trap.
Using the method of matched asymptotic expansions, we derive the velocity of
an element of vortex line as a function of the local gradient of the trap
potential, the line curvature and the angular velocity of the trap rotation.
This velocity yields small-amplitude normal modes of the vortex for 2D and
3D condensates. For an axisymmetric trap, the motion of the vortex line is a
superposition of plane-polarized standing-wave modes. In a 2D condensate,
the planar normal modes are degenerate, and their superposition can result
in helical traveling waves, which differs from a 3D condensate. Including
the effects of trap rotation allows us to find the angular velocity that
makes the vortex locally stable. For a cigar-shape condensate, the vortex
curvature makes a significant contribution to the frequency of the lowest
unstable normal mode; furthermore, additional modes with negative
frequencies appear. As a result, it is considerably more difficult to
stabilize a central vortex in a cigar-shape condensate than in a disc-shape
one. Normal modes with imaginary frequencies can occur for a nonaxisymmetric
condensate (in both 2D and 3D). In connection with recent JILA experiments,
we consider the motion of a straight vortex line in a slightly nonspherical
condensate. The vortex line changes its orientation in space at the rate
proportional to the degree of trap anisotropy and can exhibit periodic
recurrences.
\end{abstract}

\section{Introduction}

The experimental achievement of Bose-Einstein condensation in confined
alkali-atom gases~\cite{And,Brad1,Dav} has stimulated great interest in the
generation and observation of vortices in such systems~\cite{JMA,DCLZ,MZW}.
Rotating a totally anisotropic harmonic trap at an angular frequency $\Omega$
can, in principle, generate vortices; they are energetically stable for $
\Omega >\Omega _c$~\cite{BP,D,Svid98a,FCS}. There are several other ideas to
create vortices in a trapped Bose-Einstein condensate (BEC)~\cite
{DCLZ,Jack98,Bold98a,Wini99,Angl99,Mars99,Drum99,Dobr99,Jack99,Olsh98,Ruos99}.
Vortex formation in BEC was recently observed experimentally~\cite
{Matt99,Madi99,Madi00,Chev00,Ande00}.

In general, a vortex line in a trapped Bose-Einstein condensate is
nonstationary. The vortex line can move as a whole, undergo deformation of
its shape or perform oscillatory motion like helical waves~\cite
{Pita61,Svid98}. An extensive literature exists on vortex dynamics in
superfluids~\cite{Donn91}. The nonlinear Schr\"odinger equation
(Gross-Pitaevskii model)~\cite{Lund91,Rica92,Rica90,Kopl93,Kopl96} has
served to study the dynamics and reconnection of vortices, their time
evolution, and scattering interactions of superfluid vortex rings. Vortex
precession in a nonuniform light beam has recently been observed and
discussed in terms of the nonlinear Schr\"odinger equation~\cite{Kivs98}.

The dynamics of a vortex line in a spatially inhomogeneous two-dimension
(2D) condensate was considered in~\cite{RP94,Lund00}, while the problem of
curvature-driven motion of a vortex line in a homogeneous superfluid in
three dimensions (3D) was studied in~\cite{Pism91}. A normal mode with
negative frequency that corresponds to a vortex precession was found
numerically~\cite{Dodd97} and analytically for a large 3D disk-shape
BEC~\cite{Svid98a}, and for a small BEC~\cite{Linn99a}. The motion of vortex
lines and rings in Bose-Einstein condensates in harmonic traps was studied
in 2D and 3D by numerical solution of the Gross-Pitaevskii equation~\cite
{Jack99a}. Minimum energy configurations of vortices in a rotating trap were
considered in~\cite{Cast99}.

In a nonrotating trap, the vortex state has a higher energy than the
ground-state Bose condensate, so that the vortex is thermodynamically
unstable~\cite{Svid98a}. However, the vortex (with unit circulation quantum)
is dynamically stable and can decay only in the presence of dissipation.
Dissipative dynamics and the decay time of the vortex state (due to the
interaction of the vortex with the thermal cloud) in a trapped
Bose-condensed gas are discussed in~\cite{Fedi99}, where the friction
coefficient is found to be proportional to the temperature. At temperatures
relevant to current experiments, one can neglect dissipation in studying the
normal modes of the vortex because the vortex decay rate is much smaller
than the frequencies of the normal modes.

In this paper we consider the dynamics of a vortex line in a
zero-temperature condensate in the Thomas-Fermi (TF) limit, when the vortex
core radius $\xi \sim d^2/R$ is small compared to the mean oscillator length
$d$ and the mean dimension $R$ of the condensate [here, $d=(d_xd_yd_z)^{1/3}$
with $d_i=\sqrt{\hbar /M\omega _i}$ and trap frequencies $\omega _i$ ($
i=x,y,z$)]. We derive a general nonlinear equation for the motion of the
vortex that includes the effects of the trap potential, the vortex curvature
and the angular velocity of the trap rotation [see Eq.~(\ref{41}) below].
Linearization of this equation around stationary configurations gives rise
to the equation for the normal modes of the vortex line. We investigate
normal modes of the vortex in 2D and 3D condensates. For a 2D condensate
there are solutions in the form of helical waves. For nonrotating trap some
of the solutions have negative eigenfrequencies (these modes are formally
unstable); furthermore, in a nonaxisymmetric trap, some solutions can have
imaginary eigenfrequencies, implying that a straight central vortex line is
unstable with respect to finite self-induced curvature.

In a 3D condensate the spectrum of normal modes becomes discrete. For a
vortex near the $z$ axis, the number of normal modes with negative frequency
depends on the aspect ratio $R_{\perp }/R_z$. A vortex in a disc-shape
condensate ($R_z<R_{\perp }$) has only one mode with negative frequency.
However, if we change the aspect ratio to a cigar-shape condensate with $
R_{\perp }<R_z$, more modes with negative frequency appear. Thus it is more
difficult to stabilize a vortex in a cigar-shape condensate rather than in a
disc-shape one.

The plan of the paper is the following. In Sec.~II we derive a general
equation of vortex dynamics using the method of matched asymptotic
expansions. In Secs.~III and IV, we discuss the normal modes of a vortex
line for 2D and 3D condensates. In Sec.~V we investigate normal modes with
imaginary frequencies that appear for a vortex in a nonaxisymmetric
condensate. In the last section we study the motion of a straight vortex
line in a slightly nonspherical trap.

\section{General equation of the vortex dynamics}

Consider a condensate in a nonaxisymmetric trap that rotates with an angular
velocity ${\bf \Omega }$. At zero temperature in a frame rotating with the
angular velocity ${\bf \Omega }$, the trap potential $V_{{\rm tr}}$ is
time-independent, and the evolution of the condensate wave function $\Psi $
is described by the time-dependent Gross-Pitaevskii (GP) equation:

\begin{equation}
\label{1}\left( -\frac{\hbar ^2}{2M}\nabla ^2+V_{{\rm tr}}+g|\Psi |^2-\mu
(\Omega )+i\hbar {\bf \Omega }\cdot \left( {\bbox r}\times \bbox{\nabla}
\right) \right) \Psi =i\hbar \frac{\partial \Psi }{\partial t},
\end{equation}
where $V_{{\rm tr}}=\frac 12M\left( \omega _x^2x^2+\omega _y^2y^2+\omega
_z^2z^2\right) $ is the external trap potential, $g=4\pi \hbar ^2a/M>0$ is
the effective interparticle interaction strength, and $\mu (\Omega )$ is the
chemical potential in the rotating frame.

We assume that the condensate contains a $q$-fold quantized vortex with the
position vector ${\bbox r}_0(z,t)$. In this section we use the {\it method
of matched asymptotic expansions\/} to determine the vortex velocity as a
function of the local gradient of the trap potential $V_{{\rm tr}}$, the
vortex curvature $k$ and the angular velocity $\bf\Omega $, generalizing the
two-dimensional results obtained by Rubinstein and~Pismen \cite{RP94,Pism91}
to the case of a three-dimensional rotating potential. The method applies
when the external potential does not change significantly on distances
comparable with the core size $|q|\xi \ll R_{\perp }$ (this is the TF\
limit) and when the curvature is not too large ($k\ll 1/|q|\xi $); it
matches the outer asymptotic form of the solution of Eq.~(\ref{1}) in the
vortex-core region ($|{\bbox \rho }-{\bbox
\rho }_0|\lesssim |q|\xi $) with the short-distance behavior of the solution
in the region far from the vortex core ($|{\bbox \rho }-{\bbox \rho }_0|\gg
|q|\xi $).

To find the solution in the vortex-core region, one may consider Eq.~(\ref{1}%
) in a local coordinate frame centered at the point ${\bbox
r}_0$ of the vortex line that moves with the vortex velocity ${\bf V}$. In
the general case, the vortex line has a curvature $k$ that depends on the
specific element in question. We introduce a local coordinate system $
(x,y,z) $, so that the $x$ axis is directed along the vortex normal, the $y$
axis is along the binormal $\hat b$ and $z$ axis is along the tangent $\hat t
$ (see Fig.~1).


\bigskip
\centerline{\epsfxsize=0.25\textwidth\epsfysize=0.37\textwidth
\epsfbox{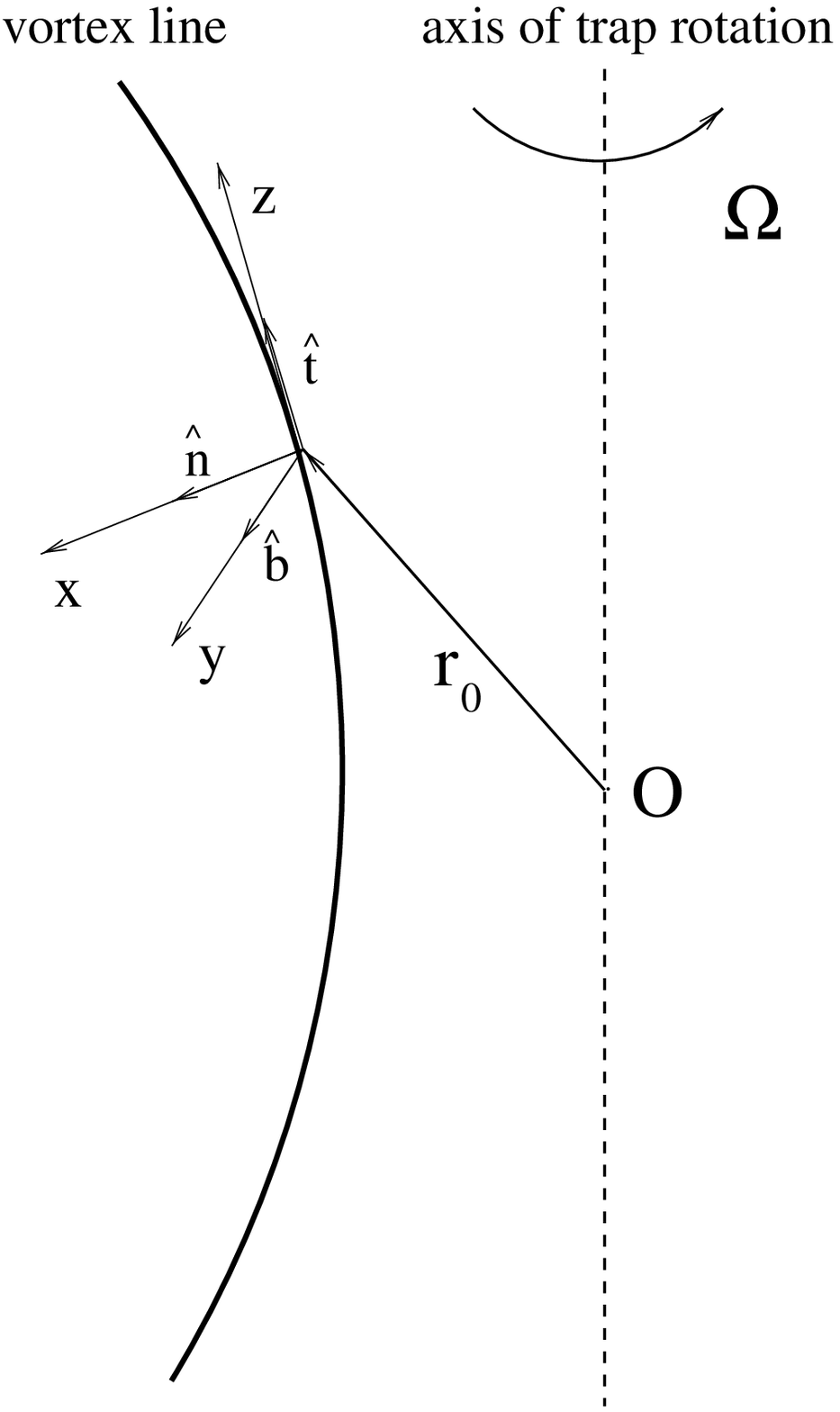}}


\begin{center}
Fig.~1. Local coordinate system associated with the vortex line.
\end{center}


The solution is assumed to be stationary in the comoving
frame and satisfies the equation (in the local coordinates):

$$
\left( -\frac{\hbar ^2}{2M}\left( \nabla ^2-k\partial _x\right) +V_{{\rm tr}
}({\bbox r}_0)+g|\Psi |^2-\mu (\Omega )+i\hbar ({\bf \Omega }\times {\bbox r}
_0)\cdot {\bbox\nabla}\right) \Psi =
$$

\begin{equation}
\label{2}=-i\hbar {\bf V}\cdot {\bbox\nabla}\Psi \enspace ,
\end{equation}
where the term $-k\partial_x$ arises from the transformation to local
coordinates.

One can remove ${\bf \Omega }$ from this equation by a shift ${\bf
V\rightarrow V-}({\bf \Omega }\times {\bbox r}_0)$. In the vortex core
region we may seek a solution in the form of an expansion in the small
parameters $\xi /R_{\perp }$ and $k\xi $:

\begin{equation}
\label{3}\Psi =\Psi _0(\rho )+\Psi _1=\left[ |\Psi _0(\rho )|-\chi
(\rho,z)\cos \phi \right] e^{iq\phi -i\eta (\rho ,z)\sin \phi },
\end{equation}
where $\Psi _0$ is the condensate wave function with $V_{{\rm tr}}$ replaced
by $V_{{\rm tr}}(\bbox{r}_0)$; it satisfies a zero-order equation

\begin{equation}
\label{4}\left( -\frac{\hbar ^2}{2M}\nabla ^2+V_{{\rm tr}}({\bbox r}
_0)+g|\Psi _0|^2-\mu (\Omega )\right) \Psi _0=0\enspace ,
\end{equation}
and $\chi $, $\eta $ characterize the perturbation in the absolute value and
phase. Physically $\Psi _0$ is the analogous wave function for a laterally
unbounded condensate with chemical potential $\mu (\Omega )-V_{{\rm tr}}(
\bbox{\rho}_0)$. The polar angle $\phi $ is measured from the direction of the
vortex normal ($\hat n\parallel \hat x$) and $\rho $ is the radial
cylindrical coordinate in the local frame.

The perturbation $\Psi _1$ obeys the following equation

\begin{equation}
\label{5}L\left( \Psi _1,\Psi _1^{*}\right) =\frac{2Mi}\hbar {\bf V}\cdot {\
\bbox \nabla }\Psi _0+\frac{2M}{\hbar ^2}\Psi _0{\bbox \rho }\cdot {\bbox
\nabla }_{\perp }V_{{\rm tr}}({\bbox r}_0)+k\partial _x\Psi _0,
\end{equation}
where

\begin{equation}
\label{6}L\left( \Psi _1,\Psi _1^{*}\right) \equiv\nabla ^2\Psi _1+\frac{2M}{
\hbar ^2}\left[ \left( \mu (\Omega )-V_{{\rm tr}}({\bbox r}_0)-2g|\Psi
_0|^2\right) \Psi _1-g\Psi _0^2\Psi _1^{*}\right]
\end{equation}
is a self-conjugate operator and ${\bbox \nabla }_{\perp }$ is the gradient
operator in a plane perpendicular to the vortex line. This equation is
linear in $\Psi _1$ and in ${\bf V}$; it contains ${\bbox \nabla }_{\perp
}V_{{\rm tr}}$ and $k$ as independent sources, so that the velocity of the
vortex line is a sum of independent contributions due to ${\bbox \nabla }
_{\perp}V_{{\rm tr}}$ and $k$. Also, the function $\Psi _0$ depends only on
the coordinates in the direction perpendicular to the vortex line;
therefore, in the dot product ${\bf V}\cdot {\bbox \nabla }\Psi _0$ only the
component of the velocity perpendicular to the vortex line is relevant. We
also assume that ${\bf V}$ has no component along the line. For simplicity,
one can assume that ${\bbox \nabla }_{\perp }V_{{\rm tr}}$ lies along $\hat n
$ and derive the vortex velocity as a sum of two independent contributions.
The final result written in vector form remains valid for arbitrary
directions of ${\bbox
\nabla }_{\perp}V_{{\rm tr}}$ and $\hat n$. Under this assumption, we have $
{\bf V}\cdot {\bbox \nabla }\Psi _0=V\partial _y\Psi _0=V\left( \sin \phi
\partial _\rho +\rho^{-1}{\cos \phi } \partial _\phi \right) \Psi _0$ in
polar coordinates. Then, writing $\Psi _1=-(\chi \cos \phi +i\eta |\Psi
_0|\sin \phi )e^{iq\phi } $ in terms of the small perturbations $\chi $ and $
\eta $, Eq.~(\ref{5}) has the form:

$$
\left( \partial _{\rho \rho }^2+\frac 1\rho \partial _\rho +\partial
_{zz}^2\right) \chi +\frac{2M}{\hbar ^2}\left( \mu (\Omega )-V_{{\rm tr}}(
\bbox{r}_0)-3 g|\Psi _0|^2\right) \chi -\frac{q^2+1}{\rho ^2}\chi -\frac{2 q
}{ \rho ^2}|\Psi _0|\eta
$$

\begin{equation}
\label{7}=\frac{2M}{\hbar ^2}|\Psi _0|\left( \frac{\hbar q}\rho V-\rho |{
\bbox
\nabla }_{\perp }V_{{\rm tr}}({\bbox r}_0)|\right) -k\partial _\rho |\Psi
_0|
\end{equation}

$$
\left( \partial _{\rho \rho }^2+\frac 1\rho \partial _\rho +\partial
_{zz}^2- \frac 1{\rho ^2}\right) \eta +\frac 2{|\Psi _0|}\left( \partial
_\rho |\Psi _0|\partial _\rho \eta +\partial _z|\Psi _0|\partial _z\eta -
\frac q{\rho ^2} \chi \right)
$$

\begin{equation}
\label{8}=-\frac{2M}\hbar V\frac{\partial _\rho |\Psi _0|}{|\Psi _0|}+\frac{
k q}\rho
\end{equation}

We can remove $V$ from these equations with the following gauge
transformation:

\begin{equation}
\label{8a}\eta =\tilde \eta -\frac M\hbar \rho V
\end{equation}
Further, for large distances $|q|\xi \ll \rho \ll R_{\perp }$, we can use $
g|\Psi _0|^2\approx g|\Psi _{TF}|^2\approx \mu (\Omega )-V_{{\rm tr}}(
\bbox{r}_0)$ and rewrite Eqs.~(\ref{7}) and (\ref{8}) as follows

\begin{equation}
\label{9}2g|\Psi _{TF}|\chi =\rho |{\bbox \nabla }_{\perp }V_{{\rm tr}}(
\bbox{\rho }_0)|+\frac{k\hbar ^2}{2M|\Psi _{TF}|}\partial _\rho |\Psi _0|,
\end{equation}

\begin{equation}
\label{10}\left( \partial _{\rho \rho }^2+\frac 1\rho \partial _\rho -\frac 1
{\rho ^2}\right) \tilde \eta -\frac \chi {|\Psi _{TF}|}\frac{2 q}{\rho ^2}=
\frac{k q}\rho;
\end{equation}

equivalently,

\begin{equation}
\label{11}\chi =\frac \rho {2g|\Psi _{TF}|}|{\bbox \nabla }_{\perp }V_{{\rm
tr} }({\bbox
r}_0)|+\frac{k\hbar ^2}{4Mg|\Psi _{TF}|^2}\partial _\rho |\Psi _0|,
\end{equation}

\begin{equation}
\label{12}\left( \rho ^2\partial _{\rho \rho }^2+\rho \partial _\rho
-1\right) \tilde \eta -\frac{q\rho |{\bbox \nabla }_{\perp }V_{{\rm tr}}(
\bbox{r}_0)|}{g|\Psi _{TF}|^2}=k q\rho +\frac{q k\hbar ^2\partial _\rho |\Psi
_0|}{ 2Mg|\Psi _{TF}|^3}\, .
\end{equation}

In Eq.~(\ref{12}), we can omit the last term, which is smaller with respect
to the term $k q\rho $ by the factor $\xi^4/\rho^4$. As a result, for $\rho
\gg |q|\xi $ the perturbations have the following asymptotic form:

\begin{equation}
\label{13}\eta \approx \frac q2\left( \frac{|{\bbox \nabla }_{\perp }V_{{\rm
tr}}(\bbox{r}_0)|}{g|\Psi _{TF}|^2}+k\right) \rho \ln \left( A\rho \right)
-\frac
M\hbar
\rho \left( {\bbox V}+{\bf \Omega }\times {\bbox r}_0\right) \cdot \hat y,
\end{equation}

\begin{equation}
\chi \approx \frac{|{\bbox \nabla }_{\perp }V_{{\rm tr}}({\bbox
r}_0)|}{2g|\Psi _{TF}|}\rho +\frac{k\hbar ^2}{4Mg|\Psi _{TF}|^2}\partial
_\rho |\Psi _0|.
\end{equation}
In terms of the phase $S$, the solution (\ref{13}) (the inner expansion in
the coordinate frame centered at the vortex line) has the form:

\begin{equation}
\label{131}S=q\phi -\frac q2\left( \frac{|{\bbox \nabla }_{\perp }V_{{\rm tr}
}({\bbox
r}_0)|}{g|\Psi _{TF}|^2}+k\right) \ln \left( A\rho \right) y+\frac M\hbar
\left( {\bbox V}+{\bf \Omega }\times {\bbox r}_0\right) \cdot {\bbox r}
\end{equation}
The parameters $A$ and ${\bbox V}$ must be determined by matching the
solution (\ref{131}) with that far from the vortex core.

To the lowest order in the small parameter $\xi /R_{\perp }$, Eq. (\ref{1})
far from the vortex core reduces to an equation for the condensate phase only

\begin{equation}
\label{14}|\Psi _{TF}|^2\nabla ^2S+{\bbox \nabla} |\Psi _{TF}|^2\cdot {
\bbox
\nabla} S-\frac M\hbar {\bf \Omega {\cdot }}\left( {\bbox r\times {\bbox
\nabla} }\right) |\Psi _{TF}|^2=0,
\end{equation}
where $\Psi =|\Psi |e^{iS}$. In the frame rotating with the trap and for $
{\bf \Omega } = \Omega \hat z$, the phase has the form

\begin{equation}
\label{132}S=S_0-\frac M\hbar \frac{(\omega _x^2-\omega _y^2)}{(\omega
_x^2+\omega _y^2)}\,\Omega \,xy,
\end{equation}
where $S_0$ is independent of $\Omega $ \cite{SF00}. Under a shift of
coordinates ${\bbox r\rightarrow \bbox r}_0+ {\bbox r}$, we have

\begin{equation}
\label{133}S\approx S_0+\frac M\hbar \left( ({\bf \Omega }\times {\bbox
r}_0)+\frac 2{M(\omega _x^2+\omega _y^2)}\left( {\bbox \nabla }V_{{\rm tr}}(
\bbox{r}_0)\times {\bf \Omega }\right) \right) \cdot {\bbox r}\, .
\end{equation}
Comparison of (\ref{131}) and (\ref{133}) allows us to find the contribution to
the vortex velocity due to the trap rotation

\begin{equation}
\label{134}{\bf V=V}_0+\frac 2{M(\omega _x^2+\omega _y^2)}\left( {\bbox
\nabla }V_{{\rm tr}}({\bbox
r}_0)\times {\bf \Omega }\right) ,
\end{equation}
where ${\bf V}_0$ is the velocity for a nonrotating trap.

It is next necessary to find the asymptotic form of $S_0$ far from the
vortex core. This function $S_0$ satisfies the following equation (in the
shifted frame):

\begin{equation}
\label{15}|\Psi _{TF}|^2\nabla ^2S_0+{\bbox \nabla }|\Psi _{TF}|^2\cdot {\
\bbox\nabla }S_0=0.
\end{equation}
Introduce a function $\Phi $ such that

\begin{equation}
\label{16}S_{0x}=-q\left( \Phi _y+\Phi \partial _y\ln |\Psi _{TF}|^2\right),
\end{equation}

\begin{equation}
\label{17}S_{0y}=q\left( \Phi _x+\Phi \partial _x\ln |\Psi _{TF}|^2\right),
\end{equation}
where $S_{0x} = \partial_x S_0$ and $S_{0y} = \partial_y S_0$. This
representation satisfies Eq. (\ref{15}) automatically. In addition, the
condition

\begin{equation}
\label{18}\hat z\cdot {\bbox \nabla}\times({\bbox \nabla} _{\perp
}S)=S_{0yx}-S_{0xy}={\bbox \nabla} _{\perp }\cdot \left( S_{0y}\hat x-S_{0x}
\hat y\right) =2\pi q\delta^{(2)} \left( {\bbox
\rho }\right)
\end{equation}
gives an equation for $\Phi $ containing a point source at the vortex
location

\begin{equation}
\label{19}{\bbox\nabla} _{\perp }\cdot \left[ {\bbox\nabla} _{\perp }\Phi
+\Phi {\bbox\nabla} _{\perp }\ln |\Psi _{TF}|^2\right] =2\pi
\delta^{(2)}\left(\bbox{\rho }\right) .
\end{equation}
Hence

\begin{equation}
\label{20}{\bbox\nabla} _{\perp }\cdot \left[ e^{-\ln |\Psi _{TF}|}{\bbox
\nabla} _{\perp }\left( \Phi e^{\ln |\Psi _{TF}|}\right) -{\bbox\nabla}
_{\perp }\left( e^{-\ln |\Psi _{TF}|}\right) e^{\ln |\Psi _{TF}|}\Phi
\right] =2\pi \delta^{(2)}\left( {\bbox \rho }\right),
\end{equation}
or

\begin{equation}
\label{21}e^{-\ln |\Psi _{TF}|}\nabla _{\perp }^2\left( \Phi e^{\ln |\Psi
_{TF}|}\right) -\nabla _{\perp }^2\left( e^{-\ln |\Psi _{TF}|}\right) e^{\ln
|\Psi _{TF}|}\Phi =2\pi \delta ^{(2)}\left( {\bbox \rho }\right) \,.
\end{equation}
We can put $\nabla _{\perp }^2\left( e^{-\ln |\Psi _{TF}|}\right) \approx
e^{-\ln |\Psi _{TF}|}\nabla _{\perp }^2V_{{\rm tr}}\,/2g|\Psi _{TF}|^2$, so
that Eq.~(\ref{21}) becomes

\begin{equation}
\label{22}\nabla _{\perp }^2\left( \Phi e^{\ln |\Psi _{TF}|}\right) -\frac{
\nabla _{\perp }^2V_{{\rm tr}}}{2g|\Psi _{TF}|^2}\Phi e^{\ln |\Psi
_{TF}|}=2\pi \delta ^{(2)}\left( {\bbox \rho }\right) e^{\ln |\Psi _{TF}|}
\end{equation}

To find the solution, we rewrite Eq. (\ref{22}) in the local coordinate
frame associated with the vortex line, taking into account the effect of
curvature:

\begin{equation}
\label{22a}\left( \nabla _{\perp }^2-k\partial _x\right) \left( \Phi e^{\ln
|\Psi _{TF}|}\right) -\frac{\nabla _{\perp }^2V_{{\rm tr}}}{2g|\Psi _{TF}|^2}
\Phi e^{\ln |\Psi _{TF}|}=2\pi \delta ^{(2)}\left( {\bbox \rho }\right)
e^{\ln |\Psi _{TF}|},
\end{equation}
or

\begin{equation}
\label{22b}\nabla _{\perp }^2\left( \Phi e^{\ln |\Psi _{TF}|-kx/2}\right)
-\left( \frac{\nabla _{\perp }^2V_{{\rm tr}}}{2g|\Psi _{TF}|^2}+\frac{k^2}4
\right) \Phi e^{\ln |\Psi _{TF}|-kx/2}=2\pi \delta ^{(2)}\left( {\bbox \rho }
\right) e^{\ln |\Psi _{TF}|}.
\end{equation}
The solution of this inhomogeneous equation is (note that an additional
solution of the homogeneous equation does not satisfy the boundary
conditions at
large $\rho $ and should be omitted):

\begin{equation}
\label{23}\Phi =-e^{kx/2}K_0\left( \sqrt{\frac{\nabla _{\perp }^2V_{{\rm tr}
} }{2g|\Psi _{TF}|^2}+\frac{k^2}4}\,\,\rho \right) ,
\end{equation}
where $K_0$ is a modified Bessel function [for small $x$, we have $
K_0\approx -\ln \left( e^Cx/2\right) $ where $C=0.577...$ is the Euler
constant]. Further, under the logarithm we may put $\nabla _{\perp }^2V_{
{\rm tr}}\,/4g|\Psi _{TF}|^2\approx 1/R_{\perp }^2$. Hence $\Phi $ (in the
local coordinate frame centered at the vortex line) has the short-distance
form

\begin{equation}
\label{24}\Phi \approx \ln \left( \frac{e^C}{\sqrt{2}}\sqrt{\frac 1{R_{\perp
}^2}+\frac{k^2}8}\,\,\rho \right).
\end{equation}

To make the asymptotic matching, we can express the solution (\ref{131}) in
the vortex-core region in terms of $\Phi $ and then compare with the formula
(\ref{24}). Using the definitions (\ref{16}), (\ref{17}) of the function $
\Phi $, one can show that the expression (\ref{131}) (for $\Omega =0$)
corresponds to the following function $\Phi $ (in the coordinate frame
centered at the vortex line):

\begin{equation}
\label{26}\Phi \approx \left[ 1+\frac x2\left( \frac{|{\bbox \nabla }_{\perp
}V_{{\rm tr}}({\bbox
r}_0)|}{g|\Psi _{TF}|^2}+k\right) \right] \ln \left( A\rho \right) +\frac{
MV_0}{\hbar q}x\, .
\end{equation}
To verify this expression, we use $\partial _x\ln |\Psi _{TF}|^2\approx -|{\
\bbox \nabla }_{\perp }V_{{\rm tr}}({\bbox
r}_0)|/g|\Psi _{TF}|^2$ and $\partial _y\ln |\Psi _{TF}|^2\approx 0$ in the
local coordinate frame where $\Phi _x\rightarrow \left( \partial _x-k\right)
\Phi $. Substituting (\ref{26}) into (\ref{16}) and (\ref{17}), we obtain

\begin{equation}
\label{27}S_{0x}=-\frac{qy}{\rho ^2},
\end{equation}

\begin{equation}
\label{28}S_{0y}=\frac{qx}{\rho ^2}-\frac q2\left( \frac{|{\bbox \nabla }
_{\perp }V_{{\rm tr}}({\bbox
r}_0)|}{g|\Psi _{TF}|^2}+k\right) \ln \left( A\rho \right) +\frac{MV_0}\hbar
\, .
\end{equation}
Therefore, in the coordinate frame centered at the vortex line, the
remaining contribution to the phase is

\begin{equation}
\label{30}S_0=q\phi -\frac q2\left( \frac{|{\bbox \nabla }_{\perp }V_{{\rm
tr }}({\bbox
r}_0)|}{g|\Psi _{TF}|^2}+k\right) \ln \left( A\rho \right) \rho \sin \phi +
\frac M\hbar V_0\rho \sin \phi
\end{equation}
which coincides with (\ref{131}) (for $\Omega =0$). Matching (\ref{26}) and
(\ref{24}) (at $\rho \sim |q|\xi $) gives an expression for the constant $A$
(with logarithmic accuracy):

\begin{equation}
\label{31}\ln \left( Ae\right) =\ln \sqrt{\frac 1{R_{\perp }^2}+\frac{k^2}8}
,
\end{equation}
where $R_{\perp }$ is the mean transverse dimension of the condensate, and
for the velocity $V_0$.

Finally, in general vector form (in the frame rotating with the trap), the
 vortex velocity is:

\begin{equation}
\label{41}{\bf V({\bbox
r}_0)=-}\frac{q\hbar }{2M}\left( \frac{\hat t{\bf \times {\bbox\nabla }}V_{
{\rm tr}}({\bbox
r}_0)}{g|\Psi _{TF}|^2}+k\hat b\right) \ln \left( |q|\xi \sqrt{\frac 1{
R_{\perp }^2}+\frac{k^2}8}\right) +\frac{2\left( {\bbox \nabla }V_{{\rm tr}
}( {\bbox
r}_0)\times {\bf \Omega }\right) }{\Delta _{\perp }V_{{\rm tr}}({\bbox
r}_0)},
\end{equation}
where $\hat b$ is a unit vector in the direction to the vortex binormal, $
\hat t$ is a tangent vector to the vortex line and $\Delta _{\perp }$ is the
Laplacian operator in the plane perpendicular to ${\bf \Omega }$. This
formula is valid for arbitrary directions of the local gradient of the trap
potential, the normal to the vortex line and ${\bf \Omega }$. Near the
condensate boundary the denominator of the first term in this formula goes
to zero. Therefore, $\hat t\times {\bbox \nabla }V_{{\rm tr}}$ must also
vanish near the boundary, implying that $\hat t$ is parallel to$\ {\bbox
\nabla }V_{{\rm tr}}$; as a result, the vortex line obeys the boundary
condition that its axis is perpendicular to the boundary.

\section{Normal modes in two dimensions}

To understand the implications of Eq.~(\ref{41}), it is valuable to consider
first the case of a 2D condensate with ${\bf \Omega }=\Omega \hat z$ and $
\omega _z=0$ (hence no confinement in the $z$ direction). Let the vector ${
\bbox \rho }_0(z,t)=(x(z,t),y(z,t))$ describe the time-dependent position of
the vortex line during its motion, and ${\bf k}$ be the vector of the
principal curvature; then $k\hat b=\hat t\times {\bf k}$ and $d^2{\bbox \rho
}_0/ds^2= {\bf k}$, where $s$ is the length measured along the vortex line.
For small displacements of the line from the $z$ axis, we have $s\approx z$,
$\hat t \approx \hat z$ and ${\bf k\approx }d^2{\bbox \rho }_0/dz^2$. Then
using

\begin{equation}
\label{43}\hat z\times {\bbox  \nabla }V_{{\rm tr}}=-M\omega _y^2y\hat x
+M\omega _x^2x\hat y
\end{equation}
and

\begin{equation}
\label{44}k\hat b=\hat t\times {\bf k}\approx -\hat x\frac{\partial ^2y}{
\partial z^2}+\hat y\frac{\partial ^2x}{\partial z^2},
\end{equation}
we obtain the following coupled differential equations for $x(z,t)$ and $
y(z,t)$

\begin{equation}
\label{q46}\frac{\partial x}{ \partial t}{\bf =}\frac{q\hbar }{2M}\left(
\frac{2y}{R_y^2}+ \frac{\partial^2y}{\partial z^2}\right) \ln \left( |q|\xi
\sqrt{\frac 1{R_{\perp }^2}+\frac{k^2}8}\right) +\frac{4\Omega \mu }{
M(\omega _x^2+\omega _y^2)}\frac y{R_y^2} ,
\end{equation}

\begin{equation}
\label{q47}\frac{\partial y}{\partial t}{\bf =-}\frac{q\hbar }{2M}\left(
\frac{2x}{R_x^2}+\frac{\partial^2x}{\partial z^2}\right) \ln \left( |q|\xi
\sqrt{\frac 1{R_{\perp }^2}+\frac{k^2}8}\right) -\frac{4\Omega \mu }{
M(\omega _x^2+\omega _y^2)}\frac x{R_x^2}.
\end{equation}

These equations have solutions in the form of helical waves

\begin{equation}
\label{q48}x=\varepsilon _x\sin (\omega t+\kappa z+\varphi _0),\quad
y=\varepsilon _y\cos (\omega t+\kappa z+\varphi _0),
\end{equation}
with the following dispersion relation between $\omega $ and $\kappa $
(under the logarithm we take $k\approx |\kappa |$):

\begin{equation}
\label{q50}\omega =\pm \frac{q\hbar }{2MR_xR_y}\sqrt{\left( 2-\kappa
^2R_x^2- \tilde \Omega \right) \left( 2-\kappa ^2R_y^2-\tilde \Omega \right)
}\ln \left( |q|\xi \sqrt{\frac 1{R_{\perp }^2}+\frac{|\kappa |^2}8}\right) .
\end{equation}
The associated amplitudes obey the relation

\begin{equation}
\label{51}\varepsilon _y=\pm \frac{R_y}{R_x}\varepsilon _x\sqrt{\frac{
2-\kappa ^2R_x^2-\tilde \Omega }{2-\kappa ^2R_y^2-\tilde \Omega }},
\end{equation}
where

\begin{equation}
\label{52}\tilde \Omega =\frac{4MR_x^2R_y^2\Omega }{q\hbar(R_x^2+R_y^2)\ln
\left( |q|\xi \sqrt{\frac 1{R_{\perp }^2}+\frac{|\kappa |^2}8}\right) ^{-1}}
\, ,
\end{equation}
is a dimensionless rotation speed.

For a nonaxisymmetric trap (for example, $R_x>R_y$) the oscillation
frequency becomes imaginary if $\sqrt{(2-\tilde \Omega )}/R_x<|\kappa |<
\sqrt{(2-\tilde \Omega )}/R_y$, in which case the initial orientation of the
vortex line along the $z$ axis is unstable with respect to the formation of
finite curvature. For $\tilde \Omega >\tilde \Omega _m=2$, however, the
oscillation frequency is real (and positive) for any $\kappa $. Thus the
trap rotation stabilizes a vortex line that initially lies along the $z$
axis.

For a straight vortex line ($\kappa =0$), the frequency is always real. This
unstable normal mode has the most negative frequency [we choose the sign in (%
\ref{q50}) that corresponds to positive-norm solution] with

\begin{equation}
\label{53}\omega =-\frac{q\hbar }{MR_xR_y}\left( \ln \left( \frac{R_{\perp }
}{|q|\xi }\right) -\frac{4\mu \Omega }{q\hbar (\omega _x^2+\omega _y^2)}
\right) .
\end{equation}
For $q>0$ and $\Omega =0$, the vortex moves (around the $z$ axis)
counterclockwise in the positive sense. With increasing rotation frequency $
\Omega $ of the trap, the vortex velocity (as seen in the rotating frame)
decreases towards zero and vanishes at $\Omega =\Omega _m$, where the
metastable angular velocity $\Omega _m$ of trap rotation is given by

\begin{equation}
\label{54}\Omega _m=\frac{|q|\hbar (\omega _x^2+\omega _y^2)}{4\mu }\ln
\left( \frac{R_{\perp }}{|q|\xi }\right) .
\end{equation}
This value $\Omega _m$ corresponds to the angular velocity of trap rotation
at which a straight vortex line at the trap center first becomes a local
minimum of energy (for $\Omega<\Omega_m$, the central position is a local
maximum)~\cite{Svid98a}. For $\Omega >\Omega _m$ the apparent motion of the
vortex becomes clockwise. For $\kappa =0$, this straight vortex follows an
elliptic trajectory along the line $V_{{\rm tr}}=const$, as expected from
the dissipationless character of the GP equation.

For a uniform condensate ($R_x,$ $R_y\rightarrow \infty $), Eq.~(\ref{q50})
coincides with the well-known dispersion law of small oscillations of a
straight vortex line (Kelvin modes):

\begin{equation}
\label{q521}\omega =\pm \frac{q\hbar }{2M}k^2\ln (|q|\xi k).
\end{equation}

Note that one can represent the helical wave solution (\ref{q48}) as a sum
of two plane-wave solutions:
\begin{equation}
\label{c14}x_1=\varepsilon _x\cos (kz)\sin (\omega t+\varphi _0),\quad
y_1=\varepsilon _y\cos (kz)\cos (\omega t+\varphi _0),
\end{equation}

\begin{equation}
\label{c15}x_2=\varepsilon _x\sin (kz)\sin (\omega t+\varphi _0+\pi
/2),\quad \,y_2=\varepsilon _y\sin (kz)\cos (\omega t+\varphi _0+\pi /2).
\end{equation}
One can easily see that Eqs.~(\ref{c14}) and (\ref{c15}) are indeed
solutions of Eqs.~(\ref{q46})\ and (\ref{q47}) with the same dispersion law $
\omega =\omega (k)$ as the helical wave (\ref{q50}). In fact, the general
motion of the vortex line can be represented as a combination of plane-wave
solutions; helical waves are just one of the possible combinations and hence
do not represent a different set of solutions. Solutions (\ref{c14}), (\ref
{c15}) have different parity, but the same eigenfrequency (in 2D there is
degeneracy). In 3D case, the plane-wave solutions are not degenerate and,
therefore, in 3D it is impossible to construct a simple analog of helical
waves. The general vortex motion in 3D is a combination of plane waves
(plane-wave solutions in 3D exist, at least, for an axisymmetric trap) with
different numbers of nodes along the symmetry axis and hence different
frequencies.

\section{Dynamics of a vortex in three dimensions}

Let us consider small displacements of the vortex from the $z$ axis and $
{\bf \Omega }=\Omega \hat z$. The vortex curvature is proportional to the
vortex displacement, so one can put $k\approx 0$ under the logarithm in (\ref
{41}). Further, for small displacements

\begin{equation}
\label{56}\hat t\times {\bbox \nabla }V_{{\rm tr}} \approx M\left( \hat x
\left( \omega _z^2zy^{\prime }-\omega _y^2y\right) +\hat y\left( \omega
_x^2x-\omega _z^2zx^{\prime }\right) \right) ,
\end{equation}
where a prime denotes derivative with respect to $z$. Then in dimensionless
coordinates $x\rightarrow R_xx$, $y\rightarrow R_yy$, $z\rightarrow R_zz$,
Eq. (\ref{41}) becomes:

\begin{equation}
\label{58}\dot x{\bf =}\frac{q\hbar }{2MR_xR_y}\left( \frac{2\left( \beta
zy^{\prime }-y\right) }{\left( 1-z^2\right) }-\beta y^{\prime \prime
}\right) \ln \left( \frac{R_{\perp }}{|q|\xi }\right) +\frac{4\Omega \mu }{
M(\omega _x^2+\omega _y^2)}\frac y{R_xR_y},
\end{equation}

\begin{equation}
\label{59}\dot y{\bf =-}\frac{q\hbar }{2MR_xR_y}\left( \frac{2\left( \alpha
zx^{\prime }-x\right) }{\left( 1-z^2\right) }-\alpha x^{\prime \prime
}\right) \ln \left( \frac{R_{\perp }}{|q|\xi }\right) -\frac{4\Omega \mu }{
M(\omega _x^2+\omega _y^2)}\frac x{R_xR_y},
\end{equation}
where

$$
\alpha =\frac{R_x^2}{R_z^2},\qquad \beta =\frac{R_y^2}{R_z^2}
$$
are parameters characterizing the trap anisotropy. One can seek solution of
these equations in the form

$$
x=x(z)\sin (\omega t+\varphi _0),\qquad y=y(z)\cos (\omega t+\varphi _0)
$$
and obtain the following ordinary differential equations for $x(z)$, $y(z)$
and $\omega $

\begin{equation}
\label{60}\tilde \omega x{\bf =}\frac{2\left( \beta zy^{\prime }-y\right) }{
\left( 1-z^2\right) }-\beta y^{\prime \prime }+\tilde \Omega y,
\end{equation}

\begin{equation}
\label{61}\tilde \omega y{\bf =}\frac{2\left( \alpha zx^{\prime }-x\right) }{
\left( 1-z^2\right) }-\alpha x^{\prime \prime }+\tilde \Omega x,
\end{equation}
where we introduce dimensionless angular velocities

$$
\tilde \omega =\frac{2MR_xR_y\omega }{q\hbar \ln \left( \frac{R_{\perp }}{
|q|\xi }\right) },\qquad \tilde \Omega =\frac{4MR_x^2R_y^2\Omega }{q\hbar
(R_x^2+R_y^2)\ln \left( \frac{R_{\perp }}{|q|\xi }\right) }
$$

\subsection{Stationary configurations}

Consider a nonrotating trap with $\tilde \Omega =0$. In this section, we
seek stationary configurations in which the vortex line remains at rest (in
this case, the contributions to the velocity from the vortex curvature and
the trap potential compensate each other). To find the stationary
configurations we need to solve Eqs.~(\ref{60}) and (\ref{61}) with the
condition $\tilde \omega =0$. The resulting equations for $x$ and $y$
uncouple; for example, the equation for $x(z)$ has the form:

\begin{equation}
\label{q86}\left( 1-z^2\right) x^{\prime \prime }-2zx^{\prime }+\frac 2\alpha
x=0.
\end{equation}
The general solution of Eq.~(\ref{q86}) can be expressed in terms of
hypergeometric functions, but it is impossible to satisfy the boundary
conditions that $x(z)$ should be finite at $z=\pm 1$ unless $2/\alpha
=n(n+1) $, where $n$ is an integer ($n\geq 0$). In this case, the solutions
reduce to Legendre polynomials

\begin{equation}
\label{q87}x\propto P_n(z).
\end{equation}
For example, the first three physical solutions are (we ignore $n=0$ which
corresponds to $\alpha =\infty $):

\begin{equation}
\label{q88}x_1=Cz,\quad \alpha =1
\end{equation}

\begin{equation}
\label{q89}x_2=\varepsilon (1-3z^2),\quad \alpha =\frac 13
\end{equation}

\begin{equation}
\label{q90}x_3=C\left( z-\frac 53z^3\right) ,\quad \alpha =\frac 16
\end{equation}
If $2/\alpha \neq n(n+1)$, the only possible solution of Eq. (\ref{q86}) is
the trivial one with $x=0$. The equation for the $y$ coordinate has the same
solutions $y\propto P_m(z)$, if $2/\beta =m(m+1)$, and $y=0$, if $2/\beta
\neq m(m+1)$. That is, if $2/\alpha \neq n(n+1)$ and $2/\beta \neq m(m+1)$,
there are no stationary configurations of the vortex line apart from the
straight orientation along the $z$ axis.

One should note that the integer $n$ (or $m$) enumerates the solutions not
only for $\tilde \omega =0$; in particular, $n$ represents to number of
times that the vortex line (precessing with angular velocity $\tilde \omega
_n$) crosses $z$ axis. For an axisymmetric trap ($\alpha =\beta $), we can
consider $\tilde \omega _n$ as a function of $\alpha $; the function $\tilde
\omega _n$ changes sign at $\alpha =\alpha _n=2/n(n+1)$. This observation
allows us to find the number of normal modes with negative frequency at a
fixed value of anisotropy parameter $\alpha $. If $\alpha >1$ there is only
one mode with negative frequency. If $\frac 13<\alpha <1$ there are 2 such
normal modes. If $\frac 16<\alpha <\frac 13$ there are 3 modes, and so on.
If $\alpha _n<\alpha <\alpha _{n-1}$ there are $n$ normal modes with
negative frequency.

\subsection{Dynamics of a vortex in a disk-shape condensate $R_z\ll R_{\perp
}$: Investigation of unstable mode}

In the limit $\alpha $, $\beta \gg 1$ the approximate solution of Eqs.~(\ref
{60}) and (\ref{61}) that corresponds to the unstable mode is

\begin{equation}
\label{80}x=\varepsilon \left( 1+\frac{z^2}{2\alpha }\right) ,
\end{equation}

\begin{equation}
\label{81}y=\varepsilon \left( 1+\frac{z^2}{2\beta }\right) ,
\end{equation}
with the corresponding eigenvalue

\begin{equation}
\label{82}\tilde \omega =\tilde \Omega -3-\frac 1{10}\left( \frac 1\alpha +
\frac 1\beta \right) .
\end{equation}
For $\tilde \Omega =0$ the excitation energy is negative and hence formally
unstable. If the trap rotates, the solution (\ref{82}) becomes stable at $
|\Omega |\geq $ $\Omega _m$, where

\begin{equation}
\label{821}\Omega _m=\frac{|q|\hbar (\omega _x^2+\omega _y^2)}{8\mu }\left(
3+\frac 1{10}\left( \frac 1\alpha +\frac 1\beta \right) \right) \ln \left(
\frac{R_{\perp }}{|q|\xi }\right) .
\end{equation}
This expression generalizes that for the angular velocity at which a
straight vortex at the center of a thin disk-shape condensate becomes
metastable~\cite{Svid98a}, including the corrections of order $\alpha^{-1}$
and $\beta^{-1}$.

\subsection{Dynamics of a vortex in a cigar-shape condensate $R_z\gg
R_{\perp }$: Investigation of unstable modes}

In the opposite limit $\alpha $, $\beta \ll 1$, the unstable-mode solution
of Eqs.~(\ref{60}) and (\ref{61}) corresponds to exponential growth of the
vortex displacement as a function of $z$. Such a solution is possible in 3D
because the condensate is bounded along the $z$ axis.

For simplicity, consider an axisymmetric trap, so that $\alpha =\beta $. In
this case, we have only one equation because one can seek a solution in the
form $x(z)=y(z)$:

\begin{equation}
\label{83}\tilde \omega x{\bf =}\frac{2\left( \alpha zx^{\prime }-x\right) }{
\left( 1-z^2\right) }-\alpha x^{\prime \prime }+\tilde \Omega x,
\end{equation}
In the limit $\alpha \ll 1$ Eq.~(\ref{83}) has the following approximate
solution

\begin{equation}
\label{84}x=y=\varepsilon e^{|z|/\alpha },
\end{equation}

\begin{equation}
\label{85}\tilde \omega =-\frac 1\alpha +\tilde \Omega .
\end{equation}
The approximate solution (\ref{84}) has a nonanalytic behavior at $z=0$,
where a thin boundary layer appears. The actual solution differs from (\ref
{84}) in a small vicinity of $z=0$ and represents a smooth crossover from
the region $z<0$ into the region $z>0$. Equation (\ref{85}) yields the
metastable angular velocity $\Omega _m$ in an elongated cigar-shape
condensate

\begin{equation}
\label{78}\Omega _m=\frac{|q|\hbar (\omega _x^2+\omega _y^2)}{8\mu }\,\frac{
R_z^2}{R_{\perp }^2}\,\ln \left( \frac{R_{\perp }}{|q|\xi }\right) .
\end{equation}
In contrast to Eq.~(\ref{821}) for a flattened condensate, this expression
becomes very large for highly elongated trap. Consequently, it is
significantly more difficult to stabilize a vortex in a cigar-shape
condensate than in one with a disk shape.

\subsection{Numerical results for 3D}

We have used Eq.~(\ref{83}) to evaluate the eigenvalues and eigenfunctions
for an axisymmetric trap ($\alpha =\beta $). A finite trap rotation produces
only a shift of eigenvalues by $\tilde \Omega$, so we can set $\tilde \Omega
=0$. In addition, solutions of Eq.~(\ref{83}) can be classified as even or
odd functions of $z$. One can enumerate the solutions by the number $m$ of
times that the vortex line crosses $z$ axis, $m=0,1,2,\ldots $. The lowest
(most negative) eigenvalue corresponds to $m=0$.

In Fig.~2 we plot the angular velocity of the vortex precession $\tilde
\omega$ as a function of the trap anisotropy $\alpha =R_{\perp }^2/R_z^2$
for $m=0,1,2$. In appropriate limits, the numerical solution $\tilde \omega
_0$ coincides with those found analytically: $\tilde \omega _0\approx -3-
\case{1}{5}\alpha $ for $\alpha \geq 1$, and $\tilde \omega _0\approx
-1/\alpha $ for $\alpha \ll 1$. The next two solutions $\tilde \omega _1$
and $\tilde \omega _2$ are proportional to $\alpha $ for large $\alpha $ and
diverge like $-1/\alpha $ for small $\alpha $. If $\alpha >1$ only one mode
has a negative frequency (namely $\tilde\omega_0$). If $\case{1}{3}<\alpha
<1 $ there are 2 such normal modes; if $\case{1}{6}<\alpha <\case{1}{3}$
there are 3 modes, {\it etc.\/} (these numerical results coincide with those
found analytically).

The more elongated the trap, the larger the number of modes with negative
frequencies (see also \cite{Ripo99}). This conclusion represents one of our
main findings. For a disk-shape condensate, the angular velocity $\Omega _m$
for the onset of metastability is smaller than the thermodynamic critical
angular velocity $\Omega _c$, with $\Omega _m=\case{3}{5}\Omega _c$~\cite
{Svid98a}. The situation is completely different for a cigar-shape
condensate with $\alpha =R_{\perp }^2/R_z^2<0.26$, because $\Omega _m$ then
becomes larger than $\Omega _c$. For comparison, Fig.~2 includes the line $
\tilde \omega =-\tilde \Omega _c=-5$.

In Fig.~3 we plot the shape of the vortex line for the lowest (unstable)
normal mode ($m=0$) for different values of trap anisotropy $\alpha $. The
function $x(z)$ is an even function of $z$ without nodes and $\varepsilon
=x(z=0)$. In Fig.~4 and 5 we plot the shape of the vortex line for normal
modes with one and two nodes for different values of trap anisotropy $\alpha
$. In Fig.~4 $x_{\max }=|x(z=R_z)|$.


\bigskip
\centerline{\epsfxsize=0.45\textwidth\epsfysize=0.57\textwidth
\epsfbox{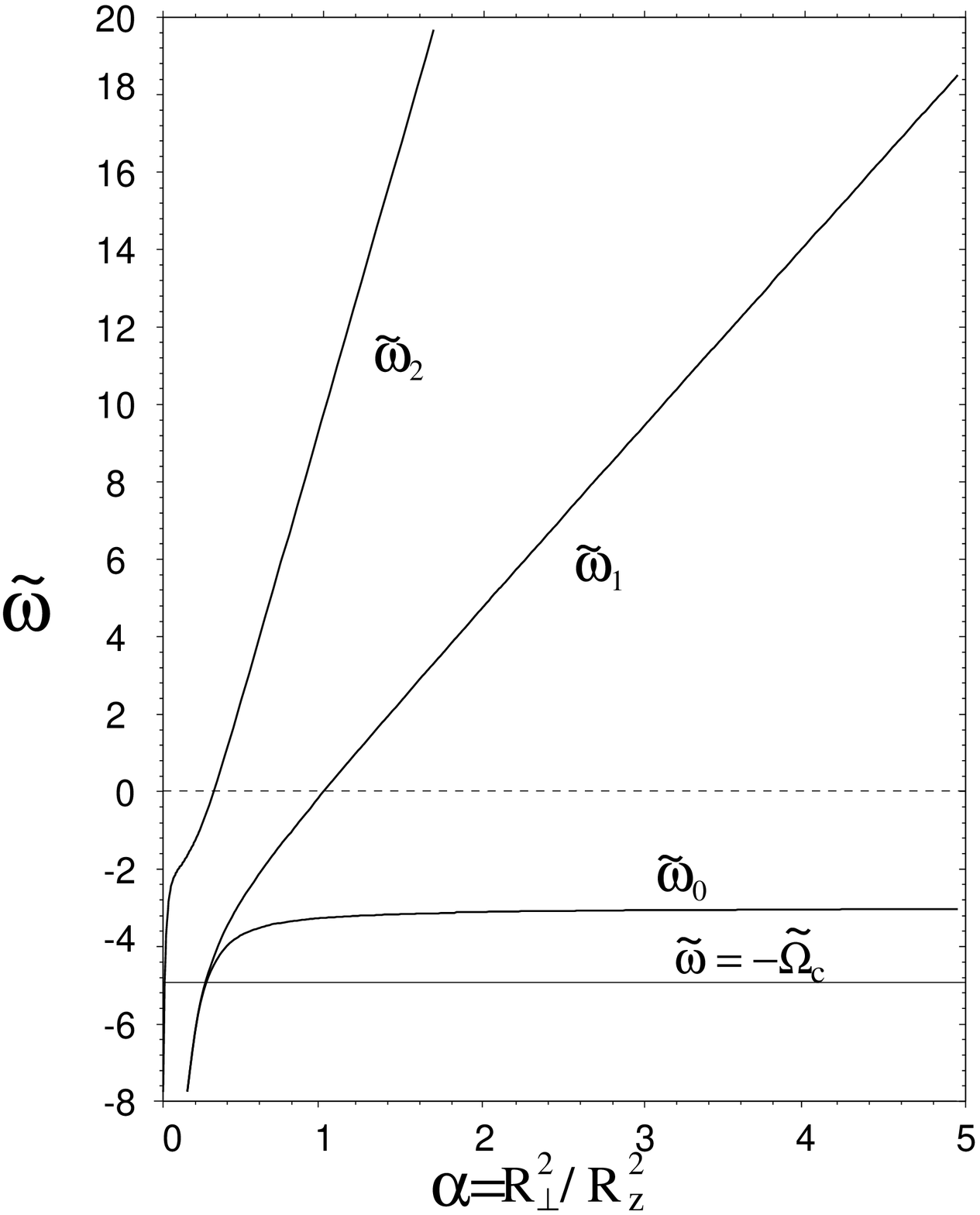}}

\nopagebreak
\begin{center}
Fig.~2. Dimentionless frequencies $\tilde \omega$ of the first three
normal modes of the vortex as a function of the trap
anisotropy $\alpha=R_{\perp}^2/R_z^2$. The lower horizontal line represents
the dimensionless thermodynamic critical angular velocity.
\end{center}




\bigskip
\centerline{\epsfxsize=0.45\textwidth\epsfysize=0.57\textwidth
\epsfbox{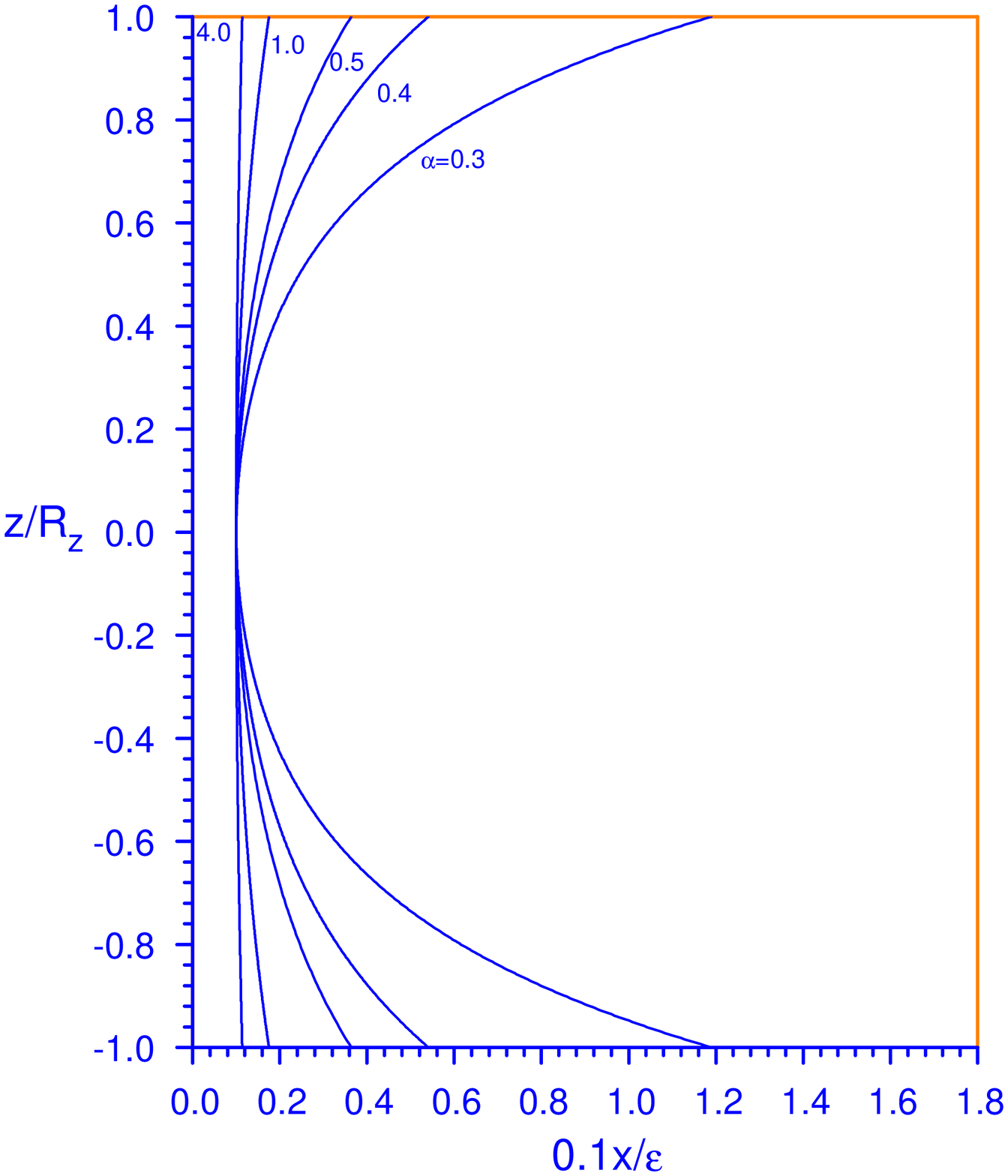}}

\nopagebreak
\begin{center}
Fig.~3. Shape of the vortex line for the normal mode with the
lowest frequency (the most unstable mode) for different values of the trap
anisotropy $\alpha$.
\end{center}



\bigskip
\centerline{\epsfxsize=0.45\textwidth\epsfysize=0.58\textwidth
\epsfbox{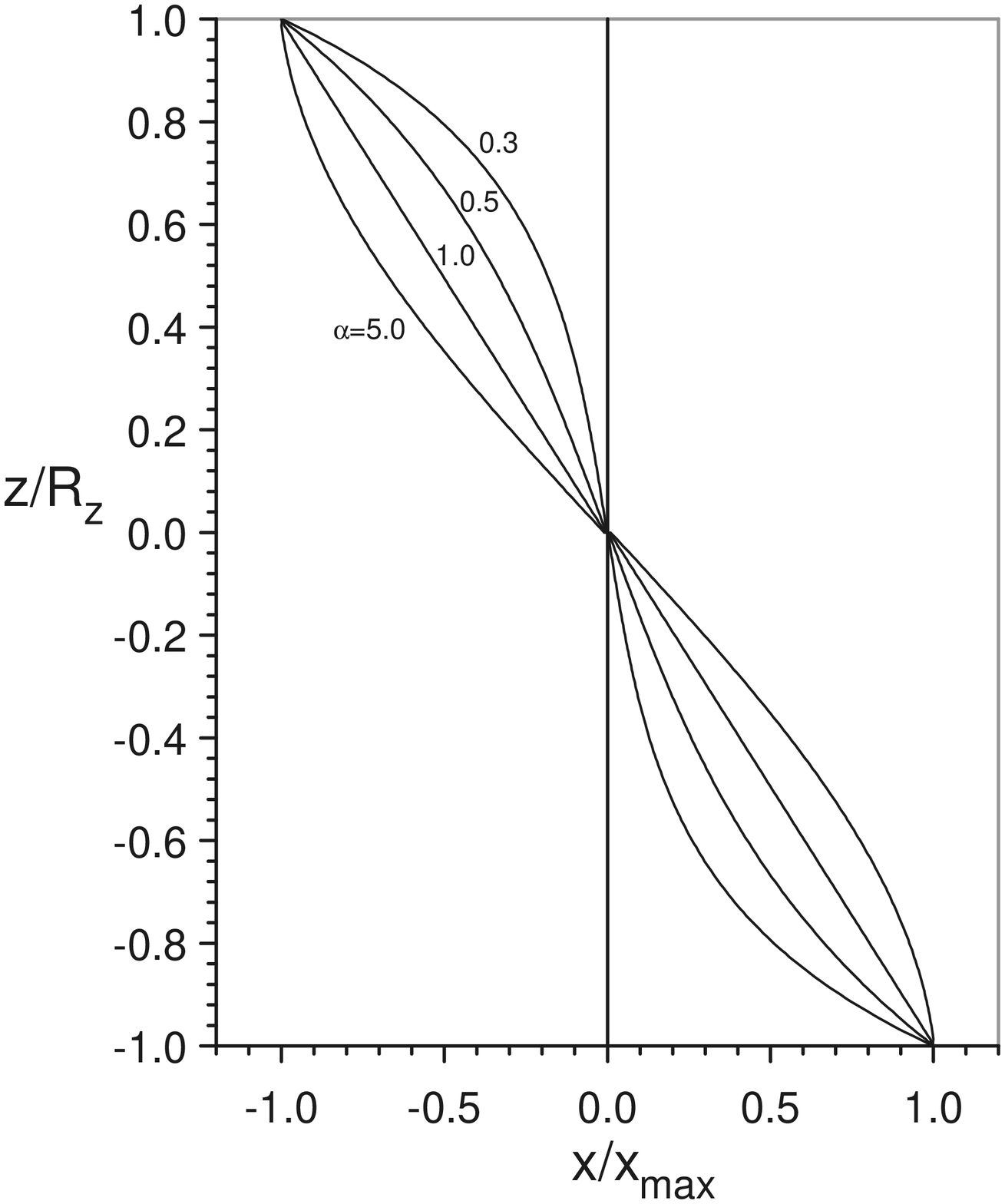}}


\begin{center}
Fig.~4. Shape of the vortex line for the first odd normal mode for different
$\alpha$.
\end{center}


\vspace{0.2cm}


\bigskip
\centerline{\epsfxsize=0.45\textwidth\epsfysize=0.58\textwidth
\epsfbox{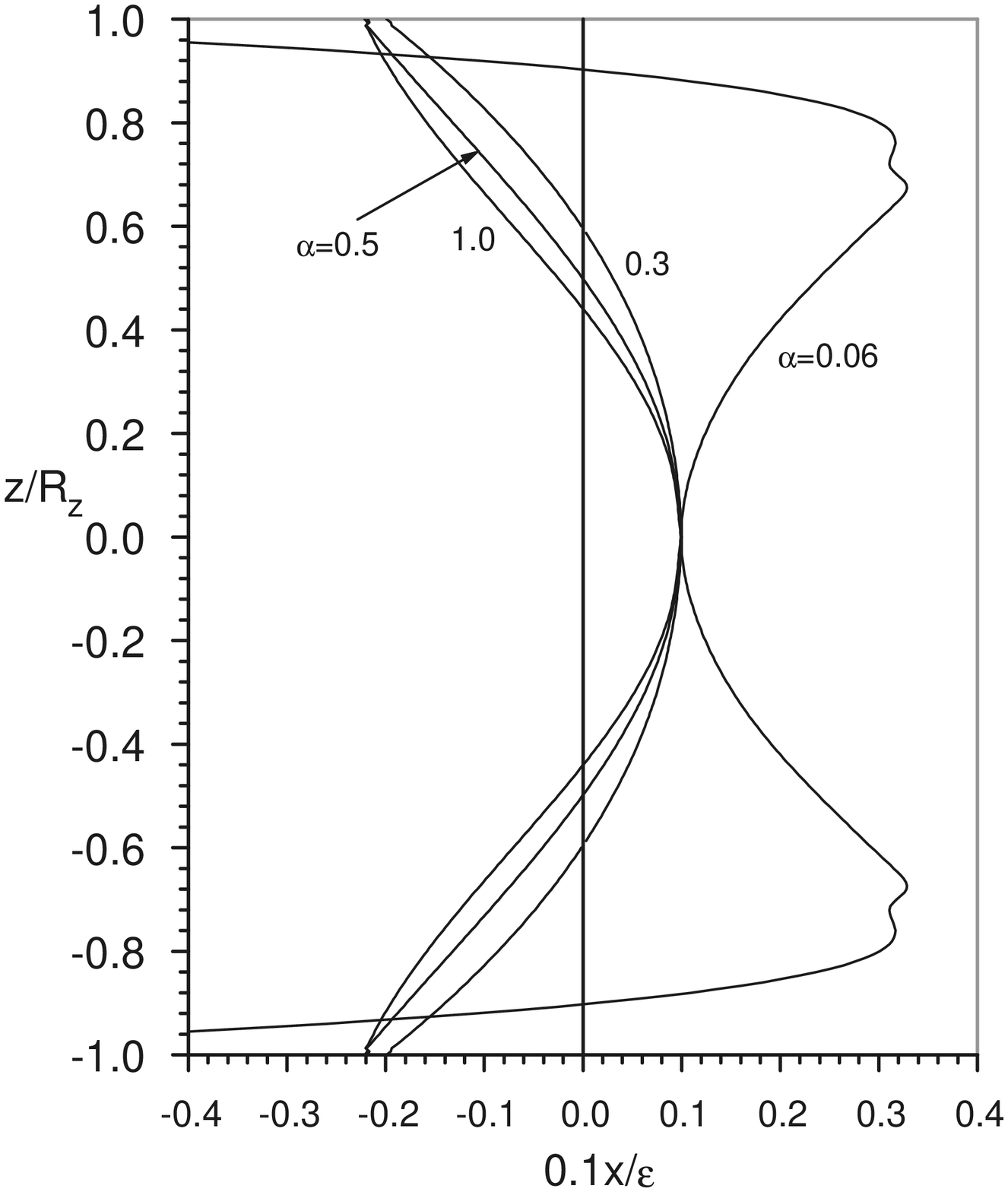}}


\begin{center}
Fig.~5. Shape of the vortex line for the second
even normal mode for different $\alpha$.
\end{center}


\section{Normal modes of a vortex with imaginary frequencies for a
nonaxisymmetric 3D condensate}

For an axisymmetric trap all small-amplitude normal modes have real
frequencies. For a nonaxisymmetric trap ($\alpha \neq \beta $), however,
Eqs.~(\ref{60}), and (\ref{61}) can have solutions with imaginary
frequencies (in certain regimes of trap anisotropy). As an example, we
recall that a spherical trap has a special normal mode with $m=1$ and $
\tilde \omega=0$. Let us now consider a nearly spherical trap (this geometry
is relevant to the JILA experiments~\cite{Corn00})

$$
|\alpha -1|\ll 1,\quad |\beta -1|\ll 1\, .
$$
One can rewrite Eqs.~(\ref{60}) and (\ref{61}) as follows:

\begin{equation}
\label{f7}\tilde \omega (1-z^2)\left(
\begin{array}{c}
x \\
y
\end{array}
\right) =\hat H_0\left(
\begin{array}{c}
x \\
y
\end{array}
\right) +\hat V\left(
\begin{array}{c}
x \\
y
\end{array}
\right) ,
\end{equation}
where

$$
\hat H_0=-\{2+\partial _z[\,(1-z^2)\,\partial _z\,]\,\}\left(
\begin{array}{cc}
0 & 1 \\
1 & 0
\end{array}
\right) ,
$$

$$
\hat V=-\partial _z\,[\,(1-z^2)\,\partial _z\,]\left(
\begin{array}{cc}
0 & \beta -1 \\
\alpha -1 & 0
\end{array}
\right) +(1-z^2)\,\tilde \Omega \left(
\begin{array}{cc}
0 & 1 \\
1 & 0
\end{array}
\right) ,
$$
and $\hat V$ is a small perturbation.

The unperturbed equation corresponds to the equation for the normal modes of
a vortex in a spherical nonrotating trap and, therefore, all
eigenfrequencies of the unperturbed equation are real. The eigenvalue $
\tilde \omega =0$ of the unperturbed equation is degenerate, and there are
two solutions that correspond to $\tilde \omega =0$:

\begin{equation}
\label{f8}\left(
\begin{array}{c}
x_1 \\
y_1
\end{array}
\right) =\left(
\begin{array}{c}
z \\
0
\end{array}
\right) ,\quad \left(
\begin{array}{c}
x_2 \\
y_2
\end{array}
\right) =\left(
\begin{array}{c}
0 \\
z
\end{array}
\right)\,.
\end{equation}
The eigenfunctions of $\hat H_0$ are Legendre polynomials, and these
eigenfunctions form a complete basis. Therefore, one can apply the usual
perturbation theory to solve Eq.~(\ref{f7}). The matrix elements are given
by:

$$
V_{11}=V_{22}=0,
$$

$$
V_{12}=\frac 43\left( \beta -1+\frac 15\tilde \Omega \right),
$$

$$
V_{21}=\frac 43\left( \alpha -1+\frac 15\tilde \Omega \right).
$$
To first order in the perturbation, the eigenfrequencies have the form

\begin{equation}
\label{f9}\tilde \omega =\pm \frac{\sqrt{V_{12}V_{21}}}{
\int_{-1}^1(1-z^2)z^2dz}=\pm 5\sqrt{\left( \alpha -1+\frac 15\tilde \Omega
\right) \left( \beta -1+\frac 15\tilde \Omega \right) }\,.
\end{equation}
If, for example, $\alpha <1<\beta $ (namely $R_x<R_z<R_y$), then the
solutions have imaginary frequencies for $\tilde \Omega<5(1-\alpha )$:

\begin{equation}
\label{f11}\tilde \omega =\pm i\gamma ,\quad \gamma =5\,\sqrt{\left(
1-\alpha -\frac 15\tilde \Omega \right) \left( \beta -1+\frac 15\tilde \Omega
\right) }>0\,.
\end{equation}
Here, the imaginary frequency means a vortex line oriented along the $z$
axis (which is the intermediate principal axis in our example) corresponds
to an unstable equilibrium. In contrast, a vortex line oriented along the
other principal axes is stable. If the angular velocity of the trap rotation
increases, however, then the solution (\ref{f9}) becomes real for $\tilde
\Omega >5(1-\alpha )$ and the vortex line along the $z$ axis becomes stable
because of the rotation.

It is straightforward to consider a general anisotropic trap (not
necessarily close to a spherical shape). The result is the following: if the
parameters $\alpha$ and $\beta $ satisfy the inequality:

\begin{equation}
\alpha <\frac 2{n(n+1)}<\beta ,
\end{equation}
where $n$ is a non-negative integer, then there is a normal-mode solution
with imaginary frequency that corresponds to the $n$th Legendre polynomial.
Moreover, if

\begin{equation}
\alpha <\frac 2{n(n+1)}<\frac 2{m(m+1)}<\beta ,
\end{equation}
then there are $(n-m+1)$ solutions with imaginary frequencies. Increasing
the external trap rotation sequentially eliminates such solutions.

\section{Motion of a straight vortex line in a slightly nonspherical 3D
condensate}

In the previous sections, we studied the motion of the vortex line for small
displacements of the vortex from equilibrium position (normal modes). In
this section we solve the general nonlinear equation of the vortex dynamics
(\ref{41}) for a slightly nonspherical trap. This problem is directly related
to a recent JILA experiment involving the evolution of an initially straight
vortex line in a nearly spherical condensate~\cite{Corn00}. In practice,
the trap
slightly deviates from the spherical shape ($R_x\ne R_y\ne R_z$).

For a strictly spherical trap, Eq.~(\ref{41}) has a solution representing a
motionless straight vortex line ($\hat t\parallel {\bf \nabla }V_{{\rm tr}}$
) that passes through the center of the trap, with the shape

\begin{equation}
\label{q1}x=\gamma _xs,\quad y=\gamma _ys,\quad z=\gamma _zs.
\end{equation}
Here $s$ is the length measured along the vortex line starting from the trap
center and $\gamma _x$, $\gamma _y$, $\gamma _z$ are the direction cosines
of the angles between the vortex line and the principal axes $x$, $y$, $z$
[so that $\gamma _x^2+\gamma _y^2+\gamma _z^2=1$, $\hat t=(\gamma _x,\gamma
_y,\gamma _z)$]. For a slightly anisotropic trap, however, the solution has
approximately the same form as (\ref{q1}), but the coefficients $\gamma _x$,
$\gamma _y$, $\gamma _z$ become time-dependent. To find a solution to first
order in the trap anisotropy, one can omit the curvature of the vortex line
and put $k=0$ in (\ref{41}). Also we should use the vortex-free condensate
density $|\Psi _{TF}|^2 $ and take $R_{\perp }$ to be equal to the value for
a spherical trap. Using the standard perturbation theory, we obtain the
following equations for the coefficients $\gamma _x$, $\gamma _y$, $\gamma
_z $ (here, we assume that there is no trap rotation):

\begin{equation}
\label{q4}
\pmatrix{\dot \gamma _x\cr\noalign{\vskip.2cm} \dot \gamma
_y\cr\noalign{\vskip.2cm}
\dot
\gamma _z}=\frac{q\hbar }{2\mu }\frac{\int_{-R}^Rs^2ds}{\int_{-R}^Rs^2\left(
1-\frac{s^2}{R^2}\right) ds}\ln \left( \frac R{|q|\xi }\right)\,
\pmatrix{
\gamma _y\gamma _z(\omega _z^2-\omega _y^2)\cr \noalign{\vskip.2cm}\gamma
_x\gamma _z(\omega _x^2-\omega _z^2)\cr\noalign{\vskip.2cm} \gamma _x\gamma
_y(\omega _y^2-\omega _x^2)}\, ,
\end{equation}
or, evaluating the integrals,

\begin{equation}
\label{q5}\dot \gamma _x=\frac{5q\hbar }{4\mu }\ln \left( \frac R{|q|\xi }
\right) (\omega _z^2-\omega _y^2)\gamma _y\gamma _z,
\end{equation}

\begin{equation}
\label{q6}\dot \gamma _y=\frac{5q\hbar }{4\mu }\ln \left( \frac R{|q|\xi }
\right) (\omega _x^2-\omega _z^2)\gamma _x\gamma _z,
\end{equation}

\begin{equation}
\label{q7}\dot \gamma _z=\frac{5q\hbar }{4\mu }\ln \left( \frac R{|q|\xi }
\right) (\omega _y^2-\omega _x^2)\gamma _x\gamma _y.
\end{equation}
Equations of this type are known in classical mechanics as Euler's equations
of rigid-body motion; they describe the evolution of the angular velocity of
free motion of a body with different principal moments of inertia as seen in
the body-fixed frame~\cite{Klep73}. Equations (\ref{q5})-(\ref{q7}) have the
following integrals of motion:

\begin{equation}
\label{q8}\gamma _x^2+\gamma _y^2+\gamma _z^2=1,
\end{equation}

\begin{equation}
\label{q9}\omega _x^2\gamma _x^2+\omega _y^2\gamma _y^2+\omega _z^2\gamma
_z^2=const.
\end{equation}
That is, the ends of the straight vortex move along trajectories that
correspond to the intersection of a sphere and an ellipsoid with principal
axes proportional to $R_x$, $R_y$, $R_z$. In fact, Eq.~(\ref{q9}) is the
equation of energy conservation during the vortex motion.

In the particular case of an axisymmetric trap (for example, $\omega
_x=\omega _y=\omega _{\perp }$), Eqs.~(\ref{q5})-(\ref{q7}) have the
following solution

\begin{equation}
\label{q10}\gamma _z=\gamma _z(0)=const,
\end{equation}

\begin{equation}
\label{q11}\gamma _x=\gamma _x(0)\cos (\omega t)+\gamma _y(0)\sin (\omega
t),
\end{equation}

\begin{equation}
\label{q12}\gamma _y=\gamma _y(0)\cos (\omega t)-\gamma _x(0)\sin (\omega
t),
\end{equation}
where

\begin{equation}
\label{q13}\omega =\frac{5q\hbar (\omega _z^2-\omega _{\perp }^2)}{4\mu }\,
\gamma _z(0)\ln \left( \frac R{|q|\xi }\right) =\frac{5q\hbar }{2M}\left(
\frac 1{R_z^2}-\frac 1{R_{\perp }^2}\right) \gamma _z(0)\ln \left( \frac R{
|q|\xi }\right) ,
\end{equation}
and [$\,\gamma _x(0),\gamma _y(0),\gamma _z(0)\,$] fixes the initial
orientation of the vortex line. The line precesses around the $z$ axis (the
axis of symmetry) at a fixed angle of inclination with respect to the $z$
axis. The frequency of this precession depends on the inclination and
vanishes if the vortex line is perpendicular to the $z$ axis with $\gamma
_z(0)=0$.

We now consider a general nonaxisymmetric trap. To be specific, we assume
that $\omega _x>\omega _y>\omega _z$ and introduce new scaled functions

\begin{equation}
\label{q14}\delta _x=\frac{5q\hbar }{4\mu }\ln \left( \frac R{|q|\xi }
\right) \sqrt{(\omega _x^2-\omega _z^2)(\omega _x^2-\omega _y^2)}\;\gamma
_x,
\end{equation}

\begin{equation}
\label{q15}\delta _y=\frac{5q\hbar }{4\mu }\ln \left( \frac R{|q|\xi }
\right) \sqrt{(\omega _y^2-\omega _z^2)(\omega _x^2-\omega _y^2)}\;\gamma
_y,
\end{equation}

\begin{equation}
\label{q16}\delta _z=\frac{5q\hbar }{4\mu }\ln \left( \frac R{|q|\xi }
\right) \sqrt{(\omega _y^2-\omega _z^2)(\omega _x^2-\omega _z^2)}\;\gamma
_z.
\end{equation}
Then one can rewrite Eqs. (\ref{q5})-(\ref{q7}) as follows:

\begin{equation}
\label{q17}\dot \delta _x=-\delta _y\delta _z,
\end{equation}

\begin{equation}
\label{q18}\dot \delta _y=\delta _x\delta _z,
\end{equation}

\begin{equation}
\label{q19}\dot \delta _z=-\delta _x\delta _y.
\end{equation}
These equations have the following property: if $\delta _x$, $\delta _y$, $
\delta _z$ is a solution of these equations, then if we change sign of any
two functions (for example, $\delta _x\rightarrow -\delta _x$, $\delta
_y\rightarrow -\delta _y$, $\delta _z\rightarrow \delta _z$), we obtain
another solution. This property can serve to construct solutions that
satisfy specific initial conditions. Equations (\ref{q17})-(\ref{q19}) have
three stationary solutions. Two of them (the vortex line parallel to the $x$
or $z $ axis) correspond to a stable equilibrium, while the third solution
(the vortex parallel to the $y$ axis) is an unstable equilibrium.

If $|\delta _x(0)|<|\delta _z(0)|$, the vortex line oscillates around $z$
axis (this is one of the equilibrium orientations), and one can express the
solution of Eqs.~(\ref{q17})-(\ref{q19}) in terms of Jacobian elliptic
functions as follows:

\begin{equation}
\label{q26}\delta _x=\sqrt{\delta _x^2(0)+\delta _y^2(0)}\,\,{\rm cn}\left(
\sqrt{ \delta _z^2(0)+\delta _y^2(0)}\,t+C,k\right) ,
\end{equation}

\begin{equation}
\label{q27}\delta _y=\pm \sqrt{\delta _x^2(0)+\delta _y^2(0)}\,\,{\rm sn}
\left( \sqrt{ \delta _z^2(0)+\delta _y^2(0)}\,t+C,k\right) ,
\end{equation}

\begin{equation}
\label{q28}\delta _z=\pm \sqrt{\delta _z^2(0)+\delta _y^2(0)}\,\,{\rm dn}
\left( \sqrt{ \delta _z^2(0)+\delta _y^2(0)}\,t+C,k\right) ,
\end{equation}
where the modulus $k=\sqrt{\delta _x^2(0)+\delta _y^2(0)}/\sqrt{\delta
_z^2(0)+\delta _y^2(0)}$ is less than one, and $C$ is a constant that must
be chosen to satisfy the initial conditions. The solution is a periodic
function of time with the period

\begin{equation}
\label{q29}T=\frac 4{\sqrt{\delta _z^2(0)+\delta _y^2(0)}}\int_0^{\pi /2}
\frac{d\varphi }{\sqrt{1-k^2\sin ^2\varphi }}=\frac{4 }{\sqrt{\delta
_z^2(0)+\delta _y^2(0)}}\,K(k) ,
\end{equation}
where $K(k)$ is the complete elliptical integral of the first kind. The
solution (\ref{q26})-(\ref{q28}) represents a superposition of a nonuniform
circular motion in a plane perpendicular to the $z $ axis and oscillations
along the $z $ axis: $\delta _z$ oscillates within the following segment:

\begin{equation}
\label{q291}\sqrt{\delta _z^2(0)-\delta _x^2(0)}\leq |\delta _z|\leq \sqrt{
\delta _z^2(0)+\delta _y^2(0)}
\end{equation}

If $|\delta _x(0)|>|\delta _z(0)|$ (the modulus $k$ is greater than one),
the vortex line oscillates around $x$ axis. In this case one can use
reciprocal modulus transformation ($k\cdot {\rm sn}(u,k)={\rm sn}(ku,1/k)$, $
{\rm cn}(u,k)={\rm dn}(ku,1/k)$, ${\rm dn}(u,k)={\rm cn}(ku,1/k)$) and
rewrite the solution (\ref{q26})-(\ref{q28}) as follows

\begin{equation}
\label{q30}\delta _x=\pm \sqrt{\delta _x^2(0)+\delta _y^2(0)}\,\,{\rm dn}
\left( \sqrt{\delta _x^2(0)+\delta _y^2(0)}\,\,t+\tilde C,1/k\right) ,
\end{equation}

\begin{equation}
\label{q31}\delta _y=\pm \sqrt{\delta _z^2(0)+\delta _y^2(0)}\,\,{\rm sn}
\left( \sqrt{ \delta _x^2(0)+\delta _y^2(0)}\,\,t+\tilde C,1/k\right) ,
\end{equation}

\begin{equation}
\label{q32}\delta _z=\sqrt{\delta _z^2(0)+\delta _y^2(0)}\,\,{\rm cn}\left(
\sqrt{ \delta _x^2(0)+\delta _y^2(0)}\,\,t+\tilde C,1/k\right).
\end{equation}
The solution is a periodic function of time with the period

\begin{equation}
\label{q33}T=\frac 4{\sqrt{\delta _z^2(0)+\delta _y^2(0)}}\int_0^{\pi /2}
\frac{d\varphi }{\sqrt{k^2-\sin ^2\varphi }}=\frac 4{\sqrt{\delta
_x^2(0)+\delta _y^2(0)}}\,{K\left( 1/k\right) }.
\end{equation}
If $|\delta _x(0)|=|\delta _z(0)|$ ($k=1$), the solution reduces to

\begin{equation}
\label{q34}\delta _x=\pm \delta _z=\frac{\sqrt{\delta _x^2(0)+\delta _y^2(0)}
}{\cosh \left( \sqrt{\delta _x^2(0)+\delta _y^2(0)}\,t+C\right) },
\end{equation}

\begin{equation}
\label{q35}\delta _y=\pm \sqrt{\delta _x^2(0)+\delta _y^2(0)}\,\tanh \left(
\sqrt{\delta _x^2(0)+\delta _y^2(0)}\,t+C\right) ;
\end{equation}
during the motion, the vortex remains on a plane through the $y$ axis,
oriented along $\delta _x=\pm \delta _z$ (see Fig.~6); it eventually lines
up along the $y$ axis (which is a direction of unstable equilibrium for the
geometry $\omega _x>\omega _y>\omega _z$ that we are considering).

Finally, one can rewrite these solutions directly in terms of the parameters
$\gamma _x$, $\gamma _y$, $\gamma _z$ that describe the orientation of the
vortex line. For example, instead of (\ref{q26})-(\ref{q29}), we have

\begin{equation}
\label{q36}\gamma _x=\sqrt{\gamma _x^2(0)+\frac{(\omega _y^2-\omega _z^2)}{
(\omega _x^2-\omega _z^2)}\gamma _y^2(0)}\cdot {\rm cn}\left( \omega
t+C,k\right) ,
\end{equation}

\begin{equation}
\label{q37}\gamma _y=\pm \sqrt{\gamma _y^2(0)+\frac{(\omega _x^2-\omega
_z^2) }{(\omega _y^2-\omega _z^2)}\gamma _y^2(0)}\cdot {\rm sn}\left( \omega
t+C,k\right) ,
\end{equation}

\begin{equation}
\label{q38}\gamma _z=\pm \sqrt{\gamma _z^2(0)+\frac{(\omega _x^2-\omega
_y^2) }{(\omega _x^2-\omega _z^2)}\gamma _y^2(0)}\cdot {\rm dn}\left( \omega
t+C,k\right) ,
\end{equation}
where

\begin{equation}
\label{q39}k=\sqrt{\frac{(\omega _x^2-\omega _y^2)}{(\omega _y^2-\omega
_z^2) }\frac{\left[ (\omega _x^2-\omega _z^2)\gamma _x^2\left( 0\right)
+(\omega _y^2-\omega _z^2)\gamma _y^2\left( 0\right) \right] }{\left[
(\omega _x^2-\omega _z^2)\gamma _z^2\left( 0\right) +(\omega _x^2-\omega
_y^2)\gamma _y^2\left( 0\right) \right] }},
\end{equation}

$$
\omega =\frac{5q\hbar \sqrt{\omega _y^2-\omega _z^2}}{4\mu }\sqrt{(\omega
_x^2-\omega _z^2)\gamma _z^2\left( 0\right) +(\omega _x^2-\omega _y^2)\gamma
_y^2\left( 0\right) }\;\ln \left( \frac R{|q|\xi }\right)
$$

\begin{equation}
\label{q390}=\frac{5q\hbar }{2M}\sqrt{\frac 1{R_y^2}-\frac 1{R_z^2}}\sqrt{
\left( \frac 1{R_x^2}-\frac 1{R_z^2}\right) \gamma _z^2\left( 0\right)
+\left( \frac 1{R_x^2}-\frac 1{R_y^2}\right) \gamma _y^2\left( 0\right) }
\;\ln \left( \frac R{|q|\xi }\right) .
\end{equation}
The period of the motion is given by

\begin{equation}
\label{q40}T=\frac 4\omega \int_0^{\pi /2}\frac{d\varphi }{\sqrt{1-k^2\sin
^2\varphi }}=\frac 4\omega \,K(k).
\end{equation}

For a nonaxisymmetric trap we plot trajectories of the end of the vortex
line in Fig. 6. The plot corresponds to the geometry $R_x<R_y<R_z$. There
are two stable orientations (along the $x$ and $z$ axes) and one unstable
(along
the $y$ axis). The shape of the trajectories strongly depends on the initial
orientation of the vortex.


\bigskip
\centerline{\epsfxsize=0.45\textwidth\epsfysize=0.65\textwidth
\epsfbox{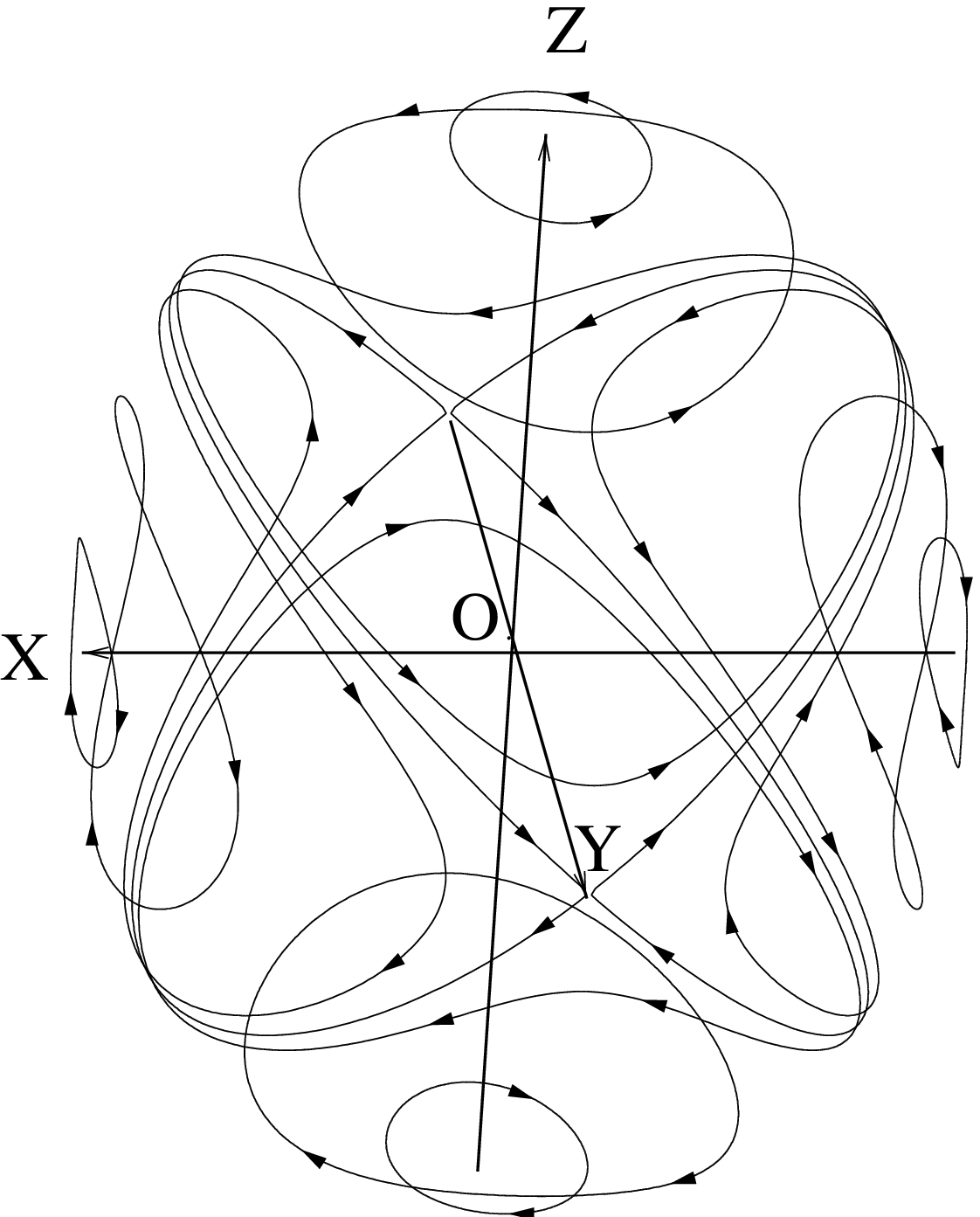}}

\vspace{0.1cm}

\begin{center}
Fig.~6. Typical trajectories of the end of a straight vortex line (that passes
through the condensate center) during its motion in a slightly nonspherical
trap with $R_x<R_y<R_z$.
\end{center}

\vspace{0.5cm}

For an axisymmetric trap, we have $\omega
_x=\omega _y$ so that $k=0$. Then, using the properties ${\rm sn}(z,0)=\sin
z $, ${\rm cn}(z,0)=\cos z$, ${\rm dn }(z,0)=1$, we can reproduce formulas (%
\ref{q10})-(\ref{q13}).

Because the motion is periodic, one can anticipate one or more revivals of
the vortex image in the recent JILA experiments. The revival time depends
on the
orientation of the vortex with respect to the symmetry axes and the
deviations from sphericity. Recently, such revivals of the vortex image (as
well as vortex precession~\cite{Ande00}) was seen by the JILA group, and their
results agree  quantitatively  with our theory~\cite{Corn00}. One
should note that our analytical results for the normal-mode frequencies are
valid with the logarithmic accuracy, namely when $\ln (R/|q|\xi )\gg 1$. It is,
however,
 straightforward to go beyond  logarithmic accuracy and obtain
the numerical correction to the logarithm (see Ref.~\cite{RP94}). This
correction
modifies our formulas for the normal-mode frequencies as follows: instead
of $\ln (R/|q|\xi )$, one should use $\ln (R/|q|\xi )+0.675=\ln
(1.96R/|q|\xi )$. In the JILA experiments, $\ln (R/\xi )\approx3.5$, and the
inclusion of the correction  improves the quantitative
agreement  with the experimental observations.

\section{Conclusions}

In this paper, we consider the dynamics of a vortex line in 2D and 3D
condensates in the TF limit. We took into account the nonuniform nature of
the system (namely the trap potential), the vortex curvature and a possible
trap rotation. We derived a general equation of vortex dynamics and
investigated various normal modes of the vortex line. For an axisymmetric
trap, all eigenvalues are real and the motion of the vortex line can be
represented as a superposition of planar normal modes with different
frequencies. In a 2D condensate, the normal modes are degenerate, and, as a
result, a superposition of planar waves can produce helical waves. In 3D,
there is no such degeneracy and there is no simple analog of the helical
waves.

An externally applied trap rotation $\Omega$ shifts the normal-mode
frequencies and makes the vortex locally stable for sufficiently large $
\Omega$. For a cigar-shape condensate, the vortex curvature has a
significant effect on the frequency of the most unstable normal mode (that
with the most negative frequency), and additional modes with (less) negative
frequencies appear. As a result, it is more difficult to stabilize central
vortex in a cigar-shape condensate than in a disc-shape one.

Normal modes with imaginary frequencies can exist for a nonaxisymmetric
condensate (both in 2D and 3D). This means that the corresponding
equilibrium orientation of the vortex is unstable. As an example of the
solution of the general nonlinear problem of vortex dynamics, we considered
the motion of a straight vortex line in a slightly nonspherical condensate.
The vortex line changes its orientation in space at a rate proportional to
the trap anisotropy.

\acknowledgments
This work was supported in part by the National Science Foundation, Grant
No. DMR 99-71518, and by Stanford University (A.A.S.).  We are grateful to
B.~Anderson, J.~Anglin, E.~Cornell and D.~Feder for valuable correspondence and
discussions.   This work benefitted from our participation in recent
workshops at
the Lorentz Center, Leiden, The Netherlands and at ECT$^*$ European Centre for
Theoretical Studies in Nuclear Physics and Related Areas, Trento, Italy; we
thank H.~Stoof and S.~Stringari for organizing these workshops and for their
hospitality.


\begin{references}
\bibitem{And}  M.~H.~Anderson {\it et al.\/}, Science~{\bf 269}, 198 (1995).

\bibitem{Brad1}  C.~C.~Bradley {\it et al.\/}, Phys.~Rev.~Lett.~{\bf 75},
1687 (1995).

\bibitem{Dav}  K.~B.~Davis {\it et al.\/}, Phys.~Rev.~Lett.~{\bf 75}, 3969
(1995).

\bibitem{JMA}  B.~Jackson, J.~F.~McCann, and C.~S.~Adams,
Phys.~Rev.~Lett.~{\bf 80}, 3903 (1998).

\bibitem{DCLZ}  R.~Dum {\it et al.\/}, Phys.~Rev.~Lett.~{\bf 80}, 2972
(1998).

\bibitem{MZW}  K.~P.~Marzlin, W.~Zhang, and E.~M.~Wright,
 Phys.~Rev.~Lett.~{\bf 79}, 4728 (1997).

\bibitem{BP}  G.~Baym and C.~J.~Pethick, Phys.~Rev.~Lett.~{\bf 76}, 6 (1996).

\bibitem{D}  F.~Dalfovo and S.~Stringari, Phys.~Rev.~A {\bf 53}, 2477 (1996).

\bibitem{Svid98a}  A.~A.~Svidzinsky and A.~L.~Fetter, Phys.~Rev.~Lett.~{\bf
84}, 5919 (2000).

\bibitem{FCS}  D.~L.~Feder, C.~W.~Clark, and B.~I.~Schneider,
Phys.~Rev.~Lett.~{\bf 82}, 4956 (1999).

\bibitem{Jack98}  B.~Jackson, J.~F.~McCann, and C.~S.~Adams,
Phys.~Rev.~Lett. {\bf 80}, 3903 (1998).

\bibitem{Bold98a}  E.~L.~Bolda and D.~F.~Walls, Phys.~Lett.~A~{\bf 246}, 32
(1998).

\bibitem{Wini99}  T.~Winiecki, J.~F.~McCann and C.~S.~Adams,
Phys.~Rev.~Lett.~{\bf 82}, 5186 (1999).

\bibitem{Angl99}  J.~R.~Anglin and W.~H.~Zurek, Phys.~Rev.~Lett.~{\bf 83},
1707 (1999).

\bibitem{Mars99}  R.~J.~ Marshall {\it et al.}, Phys.~Rev.~A {\bf 59}, 2085
(1999).

\bibitem{Drum99}  P.~D.~Drummond and J.~F.~Corney, Phys.~Rev.~A {\bf 60},
R2661 (1999).

\bibitem{Dobr99}  L.~Dobrek {\it et al.}, Phys.~Rev.~A {\bf 60}, R3381
(1999).

\bibitem{Jack99}  B.~Jackson, J.~F.~McCann, and C.~S.~Adams, Phys.~Rev.~A
{\bf 60}, 4882 (1999).

\bibitem{Olsh98}  M.~Olshanii, and M.~Naraschewski, e-print cond-mat/9811314.

\bibitem{Ruos99}  J.~Ruostekoski, B.~Kneer and W.~P.~Schleich, e-print
cond-mat/9908095.

\bibitem{Matt99}  M.~R.~Matthews {\it et al.}, Phys.~Rev.~Lett.~{\bf 83},
2498 (1999).

\bibitem{Madi99}  K.~W.~Madison {\it et al.\/}, Phys.~Rev.~Lett.~{\bf 84},
806 (2000).

\bibitem{Madi00}  K.~W.~Madison {\it et al.\/}, e-print cond-mat/0004037.

\bibitem{Chev00}  F.~Chevy, K.~W.~Madison and J.~Dalibard, e-print
cond-mat/0005221.

\bibitem{Ande00}  B.~P.~Anderson {\it et al.\/}, e-print cond-mat/0005368.

\bibitem{Pita61}  L.~P.~Pitaevskii, Zh.~Eksp.~Teor.~Fiz.~{\bf 40}, 646
(1961) [Sov.~Phys. JETP {\bf 13}, 451 (1961)].

\bibitem{Svid98}  A.~A.~Svidzinsky and A.~L.~Fetter,~Phys.~Rev.~A {\bf 58},
3168 (1998).

\bibitem{Donn91}  R.~J.~Donnelly, {\it Quantized Vortices in Helium II}
(Cambridge University Press, Cambridge, 1991).

\bibitem{Lund91}  F.~Lund, Phys.~Lett.~A {\bf 159}, 245 (1991).

\bibitem{Rica92}  S.~Rica and E.~Tirapegui, Physica~D~{\bf 61}, 246 (1992).

\bibitem{Rica90}  S.~Rica and E.~Tirapegui, Phys.~Rev.~Lett.~{\bf 64}, 878
(1990).

\bibitem{Kopl93}  J.~Koplik and H.~Levine, Phys.~Rev.~Lett. {\bf 71}, 1375
(1993).

\bibitem{Kopl96}  J.~Koplik and H.~Levine, Phys.~Rev.~Lett. {\bf 76}, 4745
(1996).

\bibitem{Kivs98}  Yu.~Kivshar {\it et al.}, Optics Comm. {\bf 152}, 198
(1998).

\bibitem{RP94}  B.~Y.~Rubinstein and L.~M.~Pismen, Physica D {\bf 78}, 1
(1994).

\bibitem{Lund00}  E.~Lundh, and P.~Ao, Phys.~Rev.~A {\bf 61}, 063612 (2000).

\bibitem{Pism91}  L.~M.~Pismen and J.~Rubinstein, Physica D {\bf 47}, 353
(1991).

\bibitem{Dodd97}  R.~J.~Dodd {\it et al.\/}, Phys.~Rev.~A {\bf 56}, 587
(1997).

\bibitem{Linn99a}  M.~Linn and A.~L.~Fetter, Phys.~Rev.~A {\bf 60}, 4910
(1999); {\bf 61}, 063603 (2000).

\bibitem{Jack99a}  B.~Jackson, J.~F.~McCann, and C.~S.~Adams, Phys.~Rev.~A
{\bf 61}, 013603 (2000).

\bibitem{Cast99}  Y.~Castin and R.~Dum, Eur.~Phys.~J.~D {\bf 7}, 399 (1999).

\bibitem{Fedi99}  P.~O.~Fedichev and G.~V.~Shlyapnikov, Phys.~Rev.~A {\bf 60}%
, R1779 (1999).

\bibitem{SF00}  A.~A.~Svidzinsky and A.~L.~Fetter, Physica B {\bf 284-288,}
21 (2000).

\bibitem{Ripo99}  J.~J.~Garc\'\i a-Ripoll and V.~M.~P\'erez-Garc\'\i a,
Phys.~Rev.~A {\bf 60}, 4864 (1999).

\bibitem{Corn00}  E.~A.~Cornell, private communication.

\bibitem{Klep73}  See, for example, D. Kleppner and R. Kolenkow, {\it An
Introduction to Mechanics\/} (McGraw-Hill, New York, 1973), Sec.~7.7.
\end{references}
\end{document}